\documentclass[aps, prd, preprint, amsfonts,
  amssymb, amsmath, showpacs, showkeys, a4paper, nofootinbib, superscriptaddress, 11pt, raggedbottom]{revtex4-1}

\usepackage{braket}
\usepackage{bm}
\usepackage{graphicx}
\usepackage{slashed}
\usepackage{subfig}
\usepackage{mathrsfs}
\usepackage{color}
\usepackage{mathtools}
\usepackage[normalem]{ulem}

\newcommand{\kket}[1]{| #1 \rangle\!\rangle}
\newcommand{\bbra}[1]{\langle\!\langle #1 |}
\newcommand{\bbrakket}[1]{\langle\!\langle #1 \rangle\!\rangle}

\def\beq{\begin{equation}}
\def\eeq{\end{equation}}
\def\beqa{\begin{eqnarray}}
\def\eeqa{\end{eqnarray}}

\begin{document}

\title{\Large Open quantum dynamics induced by light scalar fields}

\author{Clare Burrage}
\email{clare.burrage@nottingham.ac.uk}
\affiliation{School of Physics and Astronomy, University of Nottingham,\\ Nottingham NG7 2RD, United Kingdom}

\author{Christian K\"{a}ding}
\email{christian.kading@nottingham.ac.uk}
\affiliation{School of Physics and Astronomy, University of Nottingham,\\ Nottingham NG7 2RD, United Kingdom}

\author{Peter Millington}
\email{p.millington@nottingham.ac.uk}
\affiliation{School of Physics and Astronomy, University of Nottingham,\\ Nottingham NG7 2RD, United Kingdom}

\author{Ji\v{r}\'{i} Min\'{a}\v{r}}
\email{jminar@uva.nl}
\affiliation{Institute for Theoretical Physics, University of Amsterdam, \\ Science Park 904, 1098 XH Amsterdam}
\affiliation{Department of Physics, Lancaster University, \\ Lancaster, LA1 4YB, United Kingdom}
\affiliation{School of Physics and Astronomy, University of Nottingham,\\ Nottingham NG7 2RD, United Kingdom}
\affiliation{Centre for the Mathematics and Theoretical Physics of Quantum Non-Equilibrium Systems, School of Physics and Astronomy, University of Nottingham, \\ Nottingham, NG7 2RD, United Kingdom}

\begin{abstract}
We consider the impact of a weakly coupled environment comprising a light scalar field on the open dynamics of a quantum probe field, resulting in a master equation for the probe field that features corrections to the coherent dynamics, as well as decoherence and momentum diffusion. The light scalar is assumed to couple to matter either through a nonminimal coupling to gravity or, equivalently, through a Higgs portal. Motivated by applications to experiments such as atom interferometry, we assume that the probe field can be initialized, by means of external driving, in a state that is not an eigenstate of the light scalar-field--probe system, and we derive the master equation for single-particle matrix elements of the reduced density operator of a toy model. We comment on the possibilities for experimental detection and the related challenges, and highlight possible pathways for further improvements. This derivation of the master equation requires techniques of non-equilibrium quantum field theory, including the Feynman-Vernon influence functional and thermo field dynamics, used to motivate a method of Lehmann-Symanzik-Zimmermann-like reduction. In order to obtain  cutoff-independent results for the probe-field dynamics, we find that it is necessary to use a time-dependent renormalization procedure. Specifically, we show that non-Markovian effects following a quench, namely the violation of time-translational invariance due to finite-time effects, lead to a time-dependent modulation of the usual vacuum counterterms.
\end{abstract}


\maketitle


\tableofcontents

\section{Introduction}

The fundamental explanation for the evolution of the Universe on its largest scales remains a mystery, and there are many suggestions that light fields, beyond those present in the Standard Model (SM) of particle physics, could play an important role, not least in connection to the cosmological constant problem, the nature of “dark energy"~\cite{Clifton:2011jh,Joyce:2014kja} and the missing “dark matter" mass in the Universe~\cite{Hu:2000ke, Hui:2016ltb}. If such cosmologically light fields exist, they can provide an environment in which ordinary matter is coupled as an open (quantum) system.

In this work, we restrict our attention to the introduction of light \emph{scalar} fields. Light scalar fields often arise in  modifications to general relativity~\cite{Joyce:2014kja, Clifton:2011jh, Bull:2015stt} and in attempts to understand the dark energy problem, where they may be introduced directly to explain the observed accelerated expansion~\cite{Copeland:2006wr}, as well as the dark matter problem, where such fields can form condensates around galaxies and play the role of the “missing mass" in the Universe~\cite{Turner:1983he, Press:1989id, Sin:1992bg, Hu:2000ke, Goodman:2000tg, Peebles:2000yy, Arbey:2001qi, Lesgourgues:2002hk, Amendola:2005ad, Schive:2014dra, Hui:2016ltb}.\footnote{Attempts have also been made to understand galactic dynamics through modifications of gravity mediated by a light scalar field~\cite{Gessner:1992flm, Khoury:2014tka, Burrage:2016yjm, OHare:2018ayv, Burrage:2018zuj}.}

Additional scalar fields will, in general, develop interactions with the SM fields, unless there is a symmetry to prevent such couplings (or we invoke fine-tuning), and they can therefore mediate long-range fifth forces. The latter can have implications on all scales, from the cosmological to the subatomic~\cite{Burrage:2017qrf}.  However, any such forces are tightly constrained within the Solar System~\cite{ADELBERGER2009102}. In order to have avoided experimental and observational detection to date, the couplings between the light scalars and SM matter must be weak, at least locally. If the scalar field theory is nonlinear then it is not necessary to fine-tune these couplings, with the fifth force instead being suppressed dynamically in the local environment through so-called \emph{screening mechanisms}~\cite{Koyama:2015vza, Bull:2015stt}.

{
The aim of this work is to develop an approach for describing the open quantum dynamics of a \emph{quenched} subsystem of matter fields, e.g., cold atoms, coupled to an environment composed of such a light scalar field.
}

The reason for considering atoms as concrete motivation, or quantum objects more generally, such as molecules or optomechanical systems, is that they are at the forefront of precision metrology, using relatively small and cheap tabletop experiments. They are exploited to explore a wide range of phenomena. These include probing the time evolution of fundamental constants and searching for new forces~\cite{Schmiedmayer_2015,  Stadnik_2015, Berengut_2017, Stadnik_2017, Safronova_2017}, testing gravitationally induced decoherence~\cite{Bassi_2017, Minar_2016c_pub, Minar_2016b}, realizing macroscopic quantum superpositions~\cite{Frowis_2017} to probe so-called collapse models~\cite{Bassi_2013, Bahrami_2014, Li_2016, Vinante_2016, Toros_2016}, and  constraining models of dark matter and dark energy, using dielectric nanospheres levitated in laser beams~\cite{Bateman_2015, Winstone_2017}, as well as atom-interferometry searches for fifth forces mediated by light scalar fields~\cite{Burrage:2014oza, Hamilton:2015zga, Schlogel:2015uea, Jaffe:2016fsh, Elder:2016yxm, Brax_2016, Brax_2017}.

{
Working in the quantum field theory context, we develop a framework to derive quantum master equations for the matrix elements of the reduced density matrix of the probe field (our proxy for the atom/atomic system) in Fock space, which allows for quantitative and cutoff-independent predictions of the matter dynamics. We draw on a combination of techniques of non-equilibrium field theory, namely the path-integral-based Feynman-Vernon influence-functional approach~\cite{Feynman_1963, Caldeira_1983, Calzetta_1998, Breuer_2002} and the operator-based approach of thermo field dynamics~\cite{Takahasi:1974zn, Arimitsu:1985ez, Arimitsu:1985xm} (see also Ref.~\cite{Khanna}).  The latter is used to motivate a variant of Lehmann-Symanzik-Zimmermann (LSZ) reduction~\cite{Lehmann:1954rq}, which allows us to project into Fock space and extract the relevant matrix elements of the density operator. In this way, we obtain the evolution equation for the single-particle matrix element that is of interest in the low-energy limit.

Our approach has the advantage that it allows us to establish a direct connection between powerful quantum field theory techniques, including those that can account for the finite-temperature of the environment, and master equations of a form customary in atomic and condensed matter physics.

The Feynman-Vernon influence functional and similar field-theoretic approaches have been used previously, e.g., in studies of quantum Brownian motion~\cite{Hu_1992, Hu_1993}, interacting quantum field theories (both in vacuum~\cite{Koksma:2009wa} and at finite temperature~\cite{Koksma:2011dy}) and decoherence during inflation in the early Universe~\cite{Lombardo_1996, Lombardo_2005, Lombardo_2005b, Boyanovsky:2015tba, Boyanovsky:2015jen, Boyanovsky:2016exa, Boyanovsky:2018soy} (see also Refs.~\cite{Burgess_2015, Hollowood_2017, Martin_2018_JCAP, Martin_2018_JCAP_b} for the Hamiltonian approach),  as well as quarkonium suppression in heavy-ion collisions~\cite{Brambilla:2016wgg, Brambilla:2017zei}. 

A key ingredient of our approach is a novel renormalization scheme that ensures the cancellation of ultra-violet divergences for all times, also in the deeply non-Markovian regime following the quench. Here, by quench, we mean an evolution of the state of the system which is typically not an eigenstate of either the total system composed of the matter field and the light scalar or the matter field alone. Such an approach is motivated by the possibility to drive externally and initialize the matter system in an, in principle, arbitrary initial state, for instance by a sudden change of the parameters of the Hamiltonian. In the context of atom interferometry, this amounts to the initialization of the atomic state by means of external laser driving, which --- in the absence of an environment --- preserves the unitary dynamics of the atoms. Another example where quantum quenches are widely exploited is the closed (unitary) dynamics of cold atomic gases~\cite{Polkovnikov_2011_RMP,Mitra_2018_AnnRevCondMat}.

We note that extra care has to be taken when considering quantum quenches in quantum field theory. So far, most attention has been devoted to integrable theories, such as critical systems in 1+1 dimensions, which are amenable to analytical treatment \cite{Calabrese_2006_PRL, Calabrese_2007_JStatMech, Calabrese_2011_PRL, Calabrese_2012_JStatMech, Calabrese_2012_JStatMech_b}. It has been argued in the context of the sine-Gordon \cite{Bertini_2014_JStatMech}\footnote{See also Refs.~\cite{DeGrandi_2010_PRB,Iucci_2010_NJP} for further discussion of quantum quenches in sine-Gordon model.} and sinh-Gordon models \cite{Sotiriadis_2014_PhysLettB, Sotiriadis_2012_JStatMech} that, in order to avoid divergences in the postquench observables, one needs to introduce a regulating function in the quench protocol, which introduces a characteristic time-scale. While this is reminiscent of regularization, it is not equivalent as it does not introduce a single regularization scale but rather weighs the contribution of different momenta in the evaluation of the observables to make the result finite, cf.~Refs.~\cite{Sotiriadis_2014_PhysLettB, Bertini_2014_JStatMech}.

The procedure that we develop amounts to a modification of the on-shell renormalization scheme, wherein we obtain weighted integrals of the usual vacuum counterterms over the renormalization scale. This can be seen as a consequence of the Heisenberg uncertainty principle, which prevents us from determining the precise energy scale for the renormalization when the dynamics occur in a finite interval of time.  Importantly, we show that the standard on-shell renormalization is recovered in the Markovian limit, when the time-scale for the quench of the subsystem becomes long compared to its characteristic relaxation time and initial transients are neglected (see, e.g., Ref.~\cite{Blencowe_2013}). For other discussions of renormalization in open quantum systems, see, e.g., Refs.~\cite{Agon:2014uxa, Boyanovsky:2015xoa, Agon:2017oia}.

The evolution of the system following a quench should be contrasted with a scenario where a subsystem is evolving out-of-equilibrium from initial conditions which \emph{are consistent with the dynamics} of this interacting subsystem, or, in other words, where we do not assume an external impetus implementing the quench. This requires us to choose particular non-Gaussian initial states~\cite{Borsanyi:2008ar, Garny:2009ni} (see also Ref.~\cite{Agon:2017oia}). Such a choice ensures that unphysical transients and initial miscancellations of divergences are eliminated, while still using only the usual vacuum counterterms. 

Taking the chameleon model \cite{Khoury:2003rn,Khoury:2003aq} as a simple example of the light and weakly coupled fields that we have in mind, the master equation that we obtain features a coherent shift (i.e.~a correction to the unitary dynamics of the `atoms'), decoherence and momentum diffusion. These effects are, as one would anticipate, small and far beyond the reach of current atom-interferometry searches. However, the master equation for the single-particle matrix element of the reduced density operator serves as an important guide for future experimental design, illustrating, for instance, the need to maximize the momentum difference between the components of the pure initial atomic state.

The approach developed herein for describing open quantum dynamics is sufficiently general and robust that it might lead to new studies of quantum decoherence due to other long-range forces. In the context of gravitational decoherence (see, e.g., Ref.~\cite{Anastopoulos_2013, Blencowe_2013, Asprea_2019_arXiv}), it provides a new method of renormalization that should be compared with those that have been applied previously in the literature. It also allows for the analysis of the finite temperature of the light scalars on the system dynamics, as we describe in Sec. \ref{sec:Renormalization} and Sec. \ref{sec:discussion}, the inclusion of which is in principle crucial for accurate quantitative predictions. We leave this interesting application for future work.
}

We begin in Sec.~\ref{sec:lightscalar} by introducing the light scalar-field models that we have in mind, discussing, in particular, the ways in which they can be coupled to SM matter. In addition, we describe their potential screening mechanisms, focusing on the so-called chameleon mechanism, which we take as an archetype of this class of theories. We then proceed in Sec.~\ref{sec:opendynamics} to describe the derivation of the quantum master equation of the single-particle matrix element of the reduced density operator for a toy system, through which the effects of the scalar environment can be analyzed. This section includes discussions of the Feynman-Vernon influence functional and the LSZ-like reduction technique that we employ, as well as our approach to the renormalization of the loop corrections in the non-Markovian, postquench regime. We discuss the possible implications of our results for experiments in Sec.~\ref{sec:discussion}.  Our conclusions are presented in Sec.~\ref{sec:conc}.


\section{Light scalar fields}
\label{sec:lightscalar}

Two types of couplings that can lead to long-range fifth forces are commonly introduced between light, gauge-singlet scalar fields and SM matter: (i) Higgs-portal couplings and (ii) conformal (nonminimal) couplings to the Ricci scalar.

Higgs-portal couplings are often considered in the context of dark matter theories, and in extensions of the SM more generally, where they allow hidden sectors to communicate with the SM via a scalar mediator, which may itself play a role as a dark matter component. The new scalar couples directly to the Higgs field through terms of the form
\begin{equation}
\label{eq:generalportal}
\mathcal{L}\ \supset\ -\:\frac{\alpha_n}{n}\,m^{2-n} X^nH^{\dag}H\;,\qquad n\ \in\ \{1,2\}\;,
\end{equation}
where the $\alpha_n$ are dimensionless constants, $m$ is a mass scale, $H$ is the SM Higgs doublet and $X$ is the additional scalar.

On the other hand, conformal couplings are commonly considered in the context of dark energy and modified theories of gravity, although they have recently been discussed also for light scalar dark matter~\cite{Markkanen:2015xuw, Ema:2016hlw, Ema:2018ucl, Fairbairn:2018bsw, Alonso-Alvarez:2018tus}, where such nonminimal couplings can impact upon the production mechanism of the relic abundance.  In the case of modified gravity, matter fields move on geodesics of a spacetime metric $\tilde{g}_{\mu\nu} = A^2(X)g_{\mu\nu}$, where $\tilde{g}_{\mu\nu}$ is the so-called Jordan-frame metric (see, e.g., Ref.~\cite{FujiiMaeda}) and $A^2(X)$ is a coupling function.  While the Higgs-portal and nonminimal gravitational couplings would appear very different at a first glance, they are, in fact, equivalent up to second order in the scalar field $X$ (for the SM). The two descriptions are related, up to field redefinitions, by a change of conformal frame~\cite{Burrage:2018dvt}.

An additional light scalar field, which couples to matter, will mediate a fifth force, and the latter will arise at tree level so long as the coupling function $A^2(X)$ or, equivalently, the Higgs-portal interaction does not respect a $\mathbb{Z}_2$ symmetry ($X\to -X$). This can be seen by considering the geodesic equation for a matter particle moving with respect to the metric $\tilde{g}_{\mu\nu} = A^{2}(X)g_{\mu\nu}$. Alternatively, it can be seen by computing the contribution of scalar exchange to two-to-two fermion scattering in the Higgs-portal picture~\cite{Burrage:2018dvt}, wherein the coupling to fermions results from the mass mixing with the would-be SM Higgs field. The existence of long-range fifth forces is constrained by both experiments and observations, and, for a simple Yukawa force law, any such force must couple to matter approximately 5 orders of magnitude more weakly than gravity within the Solar System~\cite{Adelberger:2003zx}.

However, the scalar field can have a much more varied phenomenology in general, which can allow it to evade local tests of gravity. Specifically, whenever the scalar field theory contains nonlinearities (in the potential, in the coupling to matter, or in the kinetic structure) its properties can change depending on the environment in such a way that the fifth force it mediates can be suppressed dynamically~\cite{Joyce:2014kja}. This gives rise to the phenomenon of \emph{screening}. There is a zoo of theories that behave in this way, and this includes the Galileons~\cite{Nicolis:2008in}, chameleons~\cite{Khoury:2003rn,Khoury:2003aq} and symmetrons~\cite{Hinterbichler:2010es, Hinterbichler:2011ca} (see Refs.~\cite{Dehnen:1992rr, Damour:1994zq, Pietroni:2005pv, Olive:2007aj, Burrage:2016xzz} for similar and related models), all of which fall within the Horndeski class of scalar-tensor theories~\cite{Deffayet:2009mn, Horndeski:1974wa}.

For the sake of concreteness, we consider a specific chameleon model in this work, but the results that follow can be adapted readily to other models. 


\subsection{Einstein-frame action}
\label{sec:Chameleon}

For our purposes, it is convenient to work in the Einstein frame, wherein the gravitational sector is of canonical Einstein-Hilbert form and the action involving the conformally coupled scalar can be written in the general form~\cite{Khoury:2003rn, Khoury:2003aq}
\begin{align}
S\ &=\ \int {\rm d}^4x\; \sqrt{-\,g} \left[\frac{1}{2}\,M_{\rm Pl}^2\mathcal{R}\:-\:\frac{1}{2}g^{\mu\nu}\partial_{\mu}X\partial_{\nu}X\: -\:V(X)\right] \nonumber\\&\qquad+ \int {\rm d}^4 x\; \sqrt{-\,g}A^4(X)\tilde{\mathcal{L}}_m(\{\tilde{\phi}_i\},A^{2}(X)g_{\mu\nu})\;.
\label{eq:S chameleon}
\end{align}
Herein, $g_{\mu\nu}$ is the Einstein-frame metric, $\mathcal{R}$ is the associated Ricci scalar, $M_{\rm Pl}$ is the reduced Planck mass, $V(X)$ is the potential for the scalar $X$ and $\tilde{\mathcal{L}}_m$ is the Lagrangian density of the SM/matter fields, which we denote by the set $\{\tilde{\phi}_i\}$ and which feel the rescaled metric $\tilde{g}_{\mu\nu}=A^2(X)g_{\mu\nu}$.  Throughout this article, we use the $(-,+,+,+)$ signature convention.

In lieu of a realistic model of the probe atom, we will take the Jordan-frame matter action to be that of a single scalar field of mass $\tilde{m}^2$:
\begin{equation}
\tilde{\mathcal{L}}_m\ =\ -\:\frac{1}{2}\,\tilde{g}^{\mu\nu}\,\partial_{\mu}\tilde{\phi}\,\partial_{\nu}\tilde{\phi}\:-\:\frac{1}{2}\,\tilde{m}^2\tilde{\phi}^2\;.
\end{equation}
The Einstein-frame Lagrangian then takes the form
\begin{equation}
\mathcal{L}_m\ \equiv\ A^4(X)\tilde{\mathcal{L}}_m\ =\ -\:\frac{1}{2}\, A^2(X)\,g^{\mu\nu}\,\partial_{\mu}\tilde{\phi}\,\partial_{\nu}\tilde{\phi}\: 
-\:\frac{1}{2}\,A^4(X)\,\tilde{m}^2 \tilde{\phi}^2\;.
\end{equation}
After redefining the scalar matter field via
\begin{equation}
\phi\  \equiv\ A(X)\tilde{\phi}\;,
\end{equation}
we obtain
\begin{align}
\mathcal{L}_m \ &=\ -\:\frac{1}{2}\,g^{\mu\nu}\,\partial_{\mu}\phi\,\partial_{\nu}\phi\: -\: \frac{1}{2}\,\phi^2\,g^{\mu\nu}\,\partial_{\mu}\ln A(X)\,\partial_{\nu}\ln A(X)\nonumber\\&\qquad +\: \phi\,g^{\mu\nu}\,\partial_\mu\phi\, \partial_\nu \ln A(X)\:  
-\: \frac{1}{2}\,A^2(X)\,\tilde{m}^2 \phi^2\;.
\end{align}
Assuming that the coupling to matter is controlled by some mass scale $\mathcal{M}$ and that $X/\mathcal{M}\ll 1$, the coupling function can be expanded as
\begin{equation}
A^2(X) \ =\ a\: +\: b\, \frac{X}{\mathcal{M}}\: +\: c\, \frac{X^2}{\mathcal{M}^2}\: +\: \mathcal{O}\bigg( \frac{X^3}{\mathcal{M}^3} \bigg)\;,
\end{equation}
where $a$, $b$ and $c$ are model-specific coefficients. Keeping only operators up to dimension four (since the higher-dimension operators are suppressed by higher powers of the scale $\mathcal{M}$), the matter action then contains
\begin{equation}
\label{eqn:CanNormLagr}
\mathcal{L}_m \ =\ -\:\frac{1}{2}\,g^{\mu\nu}\,\partial_{\mu}\phi\,\partial_{\nu}\phi\:-\:\frac{1}{2}\,\bigg(a+b\,\frac{X}{\mathcal{M}} + c\,\frac{X^2}{\mathcal{M}^2} \bigg)\tilde{m}^2 \phi^2\;. 
\end{equation}

We take the following effective potential for the $X$ field:
\begin{eqnarray}
\label{eq:effpot}
V^{\rm eff}(X) \ =\ \pm\:\frac{1}{2}\,\mu^2 X^2\:+\:\frac{\lambda}{4!}\,X^4\:+\:A(X)\rho^{\rm ext}\;,
\end{eqnarray}
where $\rho^{\rm ext}$ is the covariantly conserved energy density of some external matter source (e.g.~the vacuum chamber in an atom-interferometry experiment), approximated here as a pressureless perfect fluid. The interplay between the bare mass $\mu$ (which may be tachyonic and drive spontaneous symmetry breaking), the self-interactions and the coupling to the background density will, in general, lead to a nontrivial background field configuration $\braket{X}\neq 0$,\footnote{In order to keep our notation as simple as possible, we use $\braket{X}$ to denote the \emph{classical} expectation value of the field $X$, as given by the solution to Eq. (\ref{eq:Classical}).} satisfying the classical equation of motion
\begin{equation}
\label{eq:Classical}
\Box\braket{X}\:\mp\:\mu^2\braket{X}\:-\:\frac{\lambda}{3!}\,\braket{X}^3\ =\ \frac{{\rm d}A(X)}{{\rm d}X}\bigg|_{X\,=\,\braket{X}}\rho^{\rm ext}\;.
\end{equation}

Expanding $X=\braket{X}+\chi$, and assuming that the background is constant, i.e.~$\rho^{\rm ext}$ and $\braket{X}$ are constant, we can redefine the mass of the matter scalar via 
\begin{eqnarray}
\label{eq:mrescale}
m^2\ \equiv\  \bigg(a+b\,\frac{\langle X \rangle}{\mathcal{M}} + c\,\frac{\langle X \rangle^2}{\mathcal{M}^2} \bigg)\tilde{m}^2\;,
\end{eqnarray}
and the full Lagrangian of the two-scalar system becomes
\begin{align}\label{eqn:FullLagra}
\mathcal{L}\ &=\ -\:\frac{1}{2}\,g^{\mu\nu}\,\partial_{\mu}\phi\,\partial_{\nu}\phi\: 
-\: \frac{1}{2}\,m^2 \phi^2\: 
-\:\frac{1}{2}\,\alpha_1 m \chi\phi^2\:-\:\frac{1}{4}\,\alpha_2 \chi^2\phi^2\nonumber\\&\qquad
- \:\frac{1}{2}\,g^{\mu\nu}\,\partial_{\mu}\chi\,\partial_{\nu}\chi\: -\: \frac{1}{2}\,M^2\chi^2 
\:-\: \frac{\lambda}{4!}\big(\chi^4 + 4\langle X \rangle \chi^3\big)\:-\:V(\braket{X})\;,
\end{align}
where (up to terms of order $\braket{X}^2/\mathcal{M}^2$)
\begin{subequations}
\begin{align}
\alpha_1\ &\equiv\ \frac{m}{\mathcal{M}}\bigg[\frac{b}{a}\bigg(1-\frac{b}{a}\,\frac{\braket{X}}{\mathcal{M}}\bigg)+2\,\frac{c}{a}\,\frac{\braket{X}}{\mathcal{M}}\bigg]\;,\\
\alpha_2\ &\equiv\ 2\,\frac{c}{a}\,\frac{m^2}{\mathcal{M}^2}\;,
\end{align}
\end{subequations}
and we have defined the squared mass
\begin{eqnarray}
M^2\ \equiv\ \frac{\lambda}{2}\,\braket{X}^2\: \pm\: \mu^2\:+\:c\,\frac{\rho^{\rm ext}}{\mathcal{M}^2}\;.  
\end{eqnarray}

In advance of our later analysis, we can now isolate the free and (self-)interaction parts of the action for the field $\phi$ of the ``atom'' and the fluctuations $\chi$ of the chameleon, defining
\begin{subequations}
\begin{align}
\label{eq:Sphi}
S_\phi[\phi] \ &=\ \int_x \bigg[-\:\frac{1}{2}\,g^{\mu\nu}\,\partial_{\mu}\phi\,\partial_{\nu}\phi\: -\: \frac{1}{2}\,m^2 \phi^2\bigg]\;,
\\
S_\chi[\chi] \ &=\ \int_x  \bigg[-\:\frac{1}{2}\,g^{\mu\nu}\,\partial_{\mu}\chi\,\partial_{\nu}\chi\: -\:\frac{1}{2}\, M^2 \chi^2\bigg]\;,\\
\label{eq:Schiint}
S_{\chi,{\rm int}}[\chi] \ &=\  \int_{x\,\in\,\Omega_t}\bigg[-\:\frac{\lambda}{4!}\,\big(\chi^4 + 4\braket{X}\chi^3\big)\bigg]\;,\\
\label{eq:Sint}
S_{\rm int}[\phi,\chi] \ &=\   \int_{x\,\in\,\Omega_t}
\bigg[-\:\frac{1}{2}\,\alpha_1 m \chi\phi^2\:-\:\frac{1}{4}\,\alpha_2 \chi^2\phi^2\bigg]\;,
\end{align}
\end{subequations}
where we omit the term arising from $V^{\rm eff}(\braket{X})$ [cf.~Eq.~\eqref{eq:Scontracts}] and use the shorthand notation
\begin{equation}
\int_x\ \equiv\ \int {\rm d}^4x\;.
\end{equation}
The terms in $S_{\rm int}[\phi,\chi]$ should be compared with Eq.~\eqref{eq:generalportal}, illustrating the connection with portal couplings discussed earlier.

In Eqs.~\eqref{eq:Schiint} and \eqref{eq:Sint}, we have accounted for the fact that we will later restrict the interactions to take place over a finite period of time between the quench of the initial state of the system and its subsequent measurement.
The spacetime integrals in the interaction parts of the action are therefore restricted to the hypervolume $\Omega_t=[0,t]\times\mathbb{R}^3$. We emphasize that the free parts nevertheless have support for all times (see Refs.~\cite{Millington:2012pf, Millington:2013isa}).

{
In order to make the quench manifest, one could instead extend the limits of the time integrals to the full real line and replace the coupling constants $\alpha_1$, $\alpha_2$ and $\lambda$ by time-dependent couplings, which then parametrize the effective switching on and off of the interactions by the experimental apparatus. (We might imagine preparing the system in a screened environment, before allowing it to evolve unscreened.) While we consider instantaneous switching here, we might more generally introduce a switching function that reflects the realistic preparation and quenching of the system over some finite time-scale.
}


\subsection{Chameleons}

The scalar field theory described by Eq.~\eqref{eq:S chameleon} is a chameleon model if, in the presence of a nonrelativistic matter distribution, the scalar field $X$ is stabilized at the minimum of its effective potential and the mass of small fluctuations about this minimum depends on the local energy density. It is this variation in the mass which allows the scalar fifth force to be suppressed, or screened, from local tests of gravity. Specifically, near dense sources of matter, the minimum of the effective potential $V^{\rm eff}(X)$ [see Eq.~\eqref{eq:effpot}] lies at a finite value $\braket{X}=X_{\rm min}$ and the mass becomes large, such that the fifth force is Yukawa suppressed. Instead, in the cosmological vacuum, the fluctuations are essentially massless, allowing them to propagate a long-range force of a strength comparable to gravity.  It is common to consider polynomial potentials of the form $V^{\rm eff}(X)= \Lambda^4(\Lambda/X)^n+A(X)\rho^{\rm ext}$, with integer $n$, giving a chameleon model if $n>0$ or if $n$ is a negative even integer strictly less than~$-2$.

In this work, we consider the case $n=-\,4$, in which a factor $\lambda/4!$ is usually introduced by hand in the self-coupling term, giving the effective potential
\begin{equation}
	V^{\rm eff}(X)\ =\ \frac{\lambda}{4 !} X^4\:+\:A(X)\rho^{\rm ext}\;,
	\label{eq:V bare cham}
\end{equation}
with
\begin{equation}
A(X)\ =\ e^{X/\mathcal{M}}\;.
\end{equation}
This corresponds to taking $a=1$, $b=c=2$ and $\mu=0$ in the results of the preceding subsection, giving, for instance, the coupling constants
\begin{equation}
\alpha_1\ =\ 2\,\frac{m}{\mathcal{M}}\;,\qquad
\alpha_2\ =\ \alpha_1^2\;.
\end{equation}

The quartic chameleon model, with the potential given by Eq.~\eqref{eq:V bare cham},  is currently under pressure from atom-interferometry and torsion-balance experiments, with only a small window of parameter space remaining for $\lambda$ close to one and $\mathcal{M}$ close to the Planck scale~\cite{Burrage:2017qrf}. However, we shall consider this model with screening, since it is a particularly useful prototype for evaluating the effect of the chameleon fluctuations on the dynamics of the matter fields.

\begin{figure}[t!]

\centering

\subfloat[][]{\hspace{1.5em}\includegraphics[scale=0.2]{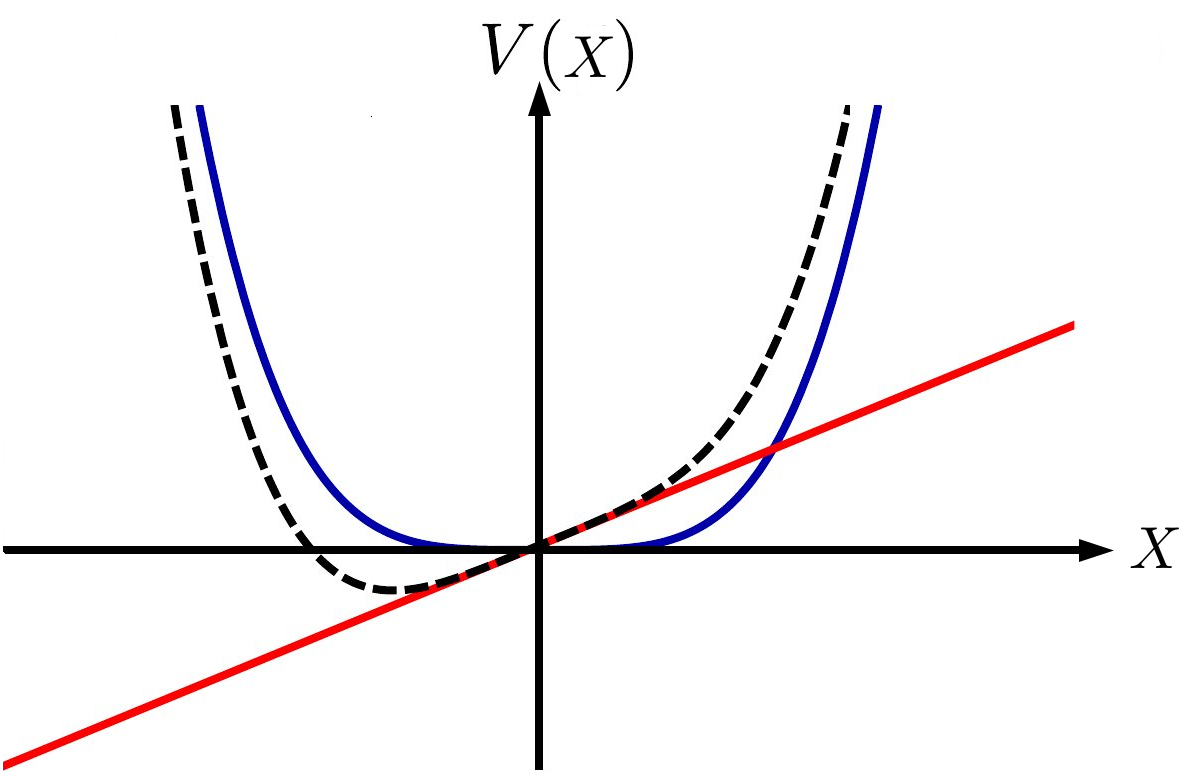}}
\qquad
\subfloat[][]{\hspace{0.5em}\includegraphics[scale=0.2]{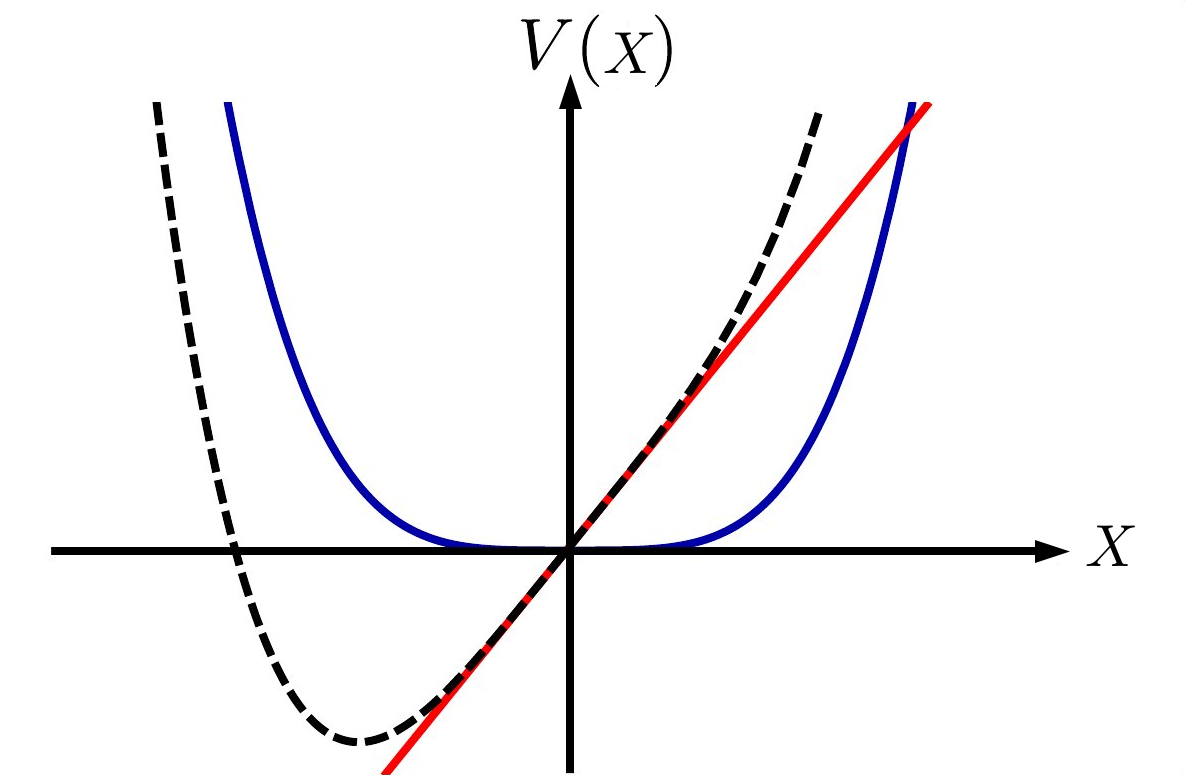}}

\caption{\label{fig:potential} The effective potential $V^{\rm eff}$, Eq. (\ref{eq:V bare cham}), at zero density (solid blue lines), the contribution from the linear coupling to matter at nonzero density (solid red lines) and the effective potential at nonzero density (black dashed lines). The panes (a) and (b) correspond to the potential in regions with lower and higher matter densities $\rho^{\rm ext}$, respectively.}

\end{figure}

In a constant-density environment, the expectation value of the chameleon field is given (to leading order in $\rho^{\rm ext}/\mathcal{M}$) by
\begin{equation}
	X_{\rm min}(\rho^{\rm ext})\ =\ -\:\bigg(\frac{6 \rho^{\rm ext} }{ \lambda \mathcal{M}}\bigg)^{1/3}\;,
\end{equation}
and the corresponding squared mass of small fluctuations around the minimum is 
\begin{equation}
	m_{\rm min}^2(\rho^{\rm ext})\ =\ \bigg( \frac{\lambda}{2}\bigg)^{1/3}\bigg(\frac{3\rho^{\rm ext}}{\mathcal{M}}\bigg)^{2/3}\;.
\end{equation}
The situation is summarized in Fig.~\ref{fig:potential}. The solid blue, solid red and black dashed lines correspond to the effective potential $V^{\rm eff}(X)$ at zero external density $\rho^{\rm ext}=0$, the contribution from the linear coupling to matter at nonzero density, and the effective potential at nonzero density, respectively. The panes (a) and (b) correspond to the potential in regions with lower and higher matter densities, respectively. In higher-density environments, the chameleon is more massive; in lower-density environments, it is less massive. 

To see how this variation in the mass of the scalar leads to a suppression of the scalar fifth force, we now consider situations where the background matter density is not uniform.  Specifically, we consider a uniform sphere of constant density $\rho^{\rm ext}$ and radius $R$ embedded in a background of lower density $\rho^{\rm bg}$.  For sufficiently large spheres, the scalar field can reach the value $X_{\rm min}$ that minimizes its effective potential at the center of the sphere. It therefore has a large mass in the interior of the sphere and does not roll from  $X_{\rm min}$ apart from in a thin shell near the surface.  It is only matter in this thin shell (whose thickness is proportional to $1/ m_{\rm min}$) that sources the chameleon fifth force in the exterior.  As only a small fraction of the mass of the sphere gives rise to the fifth force, it is much weaker than we would naively expect. This is known as the thin shell effect (see, e.g., Ref. \cite{Khoury:2003rn}).  

To see this mathematically, a sphere of radius $R$ and density $\rho^{\rm ext}$, embedded in a background of density $\rho^{\rm bg}$ has a thin-shell  at position $R_{\rm TS}$, if there exists a real solution to
\begin{equation}
1\:-\:\frac{R_{\rm TS}^2}{R^2}\ =\ \bigg(\frac{\mathcal{M}}{M_{\rm Pl}}\bigg)^2\frac{6 M_{\rm Pl}^2}{R^2 \rho^{\rm ext}}\bigg(\frac{X_{\rm min}(\rho^{\rm bg})-X_{\rm min}(\rho^{\rm ext})}{\mathcal{M}}\bigg)\;.
\end{equation}
If there is no real solution to this equation, the field cannot minimize its potential inside the sphere, and the chameleon fifth force is unscreened. The chameleon field profile around the sphere is
\begin{equation}
\braket{X}\ =\ \left\{\begin{array}{l l}
X_{\rm min}(\rho^{\rm ext})\;, &\quad 0\ <\ r\ <\ R_{\rm TS}\;,\\
X_{\rm min}(\rho^{\rm ext}) + \frac{R^2 \rho^{\rm ext}}{6 \mathcal{M}}\frac{r^3 -3 R_{\rm TS}^2 r +2 R_{\rm TS}^3}{r R^2}\;, &\quad R_{\rm TS}\ <\ r\ <\ R\;,\\
X_{\rm min}(\rho^{\rm bg}) - \frac{R^2 \rho^{\rm ext}}{3 \mathcal{M}}\left(1-\frac{R_{\rm TS}^3}{R^3}\right) \frac{R}{r}e^{-m_{\rm min}(\rho^{\rm bg})r}\;, &\quad R\ <\ r\;.
\end{array}\right.
\end{equation}
The fifth force felt by a test particle outside the sphere is $\vec{F} = -\vec{\nabla}\braket{X} / \mathcal{M}$. It is now clear that as $R_{\rm TS}$ approaches $R$, the fifth force is suppressed. In the limit where $R_{\rm TS}\rightarrow 0$, which corresponds to small (or very diffuse) spheres, the force is unsuppressed.  More details on how chameleons screen can be found in Ref.~\cite{Burrage:2017qrf}.


\section{Open dynamics}
\label{sec:opendynamics}

We now turn our attention to deriving the quantum master equation describing the open dynamics of the matter field $\phi$ in contact with an environment composed of the chameleon and its fluctuations $\chi$. 
Our strategy is to begin with the powerful path-integral techniques based on the Feynman-Vernon influence functional, before making connection with the matrix elements of the reduced density operator in Fock space by means of an LSZ-like reduction technique, motivated by the operator-based approach of thermo field dynamics. In effect, we apply a strategy familiar for deriving matrix elements of the scattering operator in the usual in-out formalism to the case of the in-in formalism of non-equilibrium field theory.


\subsection{The Feynman-Vernon influence functional}
\label{sec:Influence functional}

Our aim is to trace over the states of the conformally coupled scalar field $\chi$, so as to leave us with an open scalar system, whose reduced density matrix evolves subject to a quantum master equation. In order to derive this master equation, we could proceed directly at the operator level, by taking a partial trace of the quantum Liouville equation of the coupled scalar system. Alternatively, but equivalently, we can make use of the path-integral description provided by the Feynman-Vernon influence functional~\cite{Feynman_1963}, having the advantage that we can exploit the power of the diagrammatic expansions to which functional approaches lend themselves.

Our starting point is the reduced density operator of the scalar system
\begin{equation}
\hat{\rho}_{\phi}(t)\ =\ {\rm tr}_{\chi}\,\hat{\rho}(t)\;.
\end{equation}
We can evaluate the partial trace on the right-hand side by inserting complete sets of eigenstates of the Heisenberg-picture field operators $\hat{\chi}$ at time $t$.\footnote{Note that we have allowed for an explicit time dependence of the density operator in the Heisenberg picture in order to account for external driving of the system by the apparatus implementing the quench.} In this way, we obtain
\begin{equation}
\hat{\rho}_{\phi}(t)\ =\ \int {\rm d}\chi_t^{\pm}\; \delta(\chi_t^+-\chi_t^-)\hat{\rho}[\chi_t^{\pm};t]\;,
\end{equation}
where
\begin{equation}
\hat{\rho}[\chi^{\pm}_t;t]\ \equiv\ \braket{\chi^+_t|\hat{\rho}(t)|\chi^-_t}\;,
\end{equation}
which includes a functional delta function that sets $\chi^+=\chi^-$ at the time $t$. We have introduced the measure
\begin{equation}
\int {\rm d}\chi^{\pm}_t\ \equiv\ \int {\rm d}\chi_t^+{\rm d}\chi_t^-
\end{equation}
and suppress the spatial dependence of the field eigenstates $\ket{\chi^{\pm}_t}$ for notational convenience. Notice that we have introduced two copies of the eigenstates at the time $t$, distinguished by the superscript $\pm$; this amounts to the usual doubling of degrees of freedom needed to write a path-integral representation of the trace of an operator, giving rise to the Schwinger-Keldysh closed-time-path formalism~\cite{Schwinger:1960qe, Keldysh:1964ud} (see also Sec.~\ref{sec:TFD}). We can now take matrix elements of the reduced density operator in the basis of $\phi$ field eigenstates, defining the reduced density functional
\begin{equation}
\label{eq:reduceddensfunc}
\rho_{\phi}[\phi^{\pm}_t;t]\ \equiv\ \braket{\phi^+_t|\hat{\rho}_{\phi}(t)|\phi^-_t}\ =\ \int {\rm d}\chi_t^{\pm}\; \delta(\chi_t^+-\chi_t^-)\rho[\phi_t^{\pm},\chi_t^{\pm};t]\;,
\end{equation}
where
\begin{equation}
\rho[\phi^{\pm}_t,\chi^{\pm}_t;t]\ \equiv\ \braket{\phi^+_t,\chi^+_t|\hat{\rho}(t)|\phi^-_t,\chi^-_t}\;.
\end{equation}

\begin{figure}[t!]
	\begin{center}
		\includegraphics[scale=0.15]{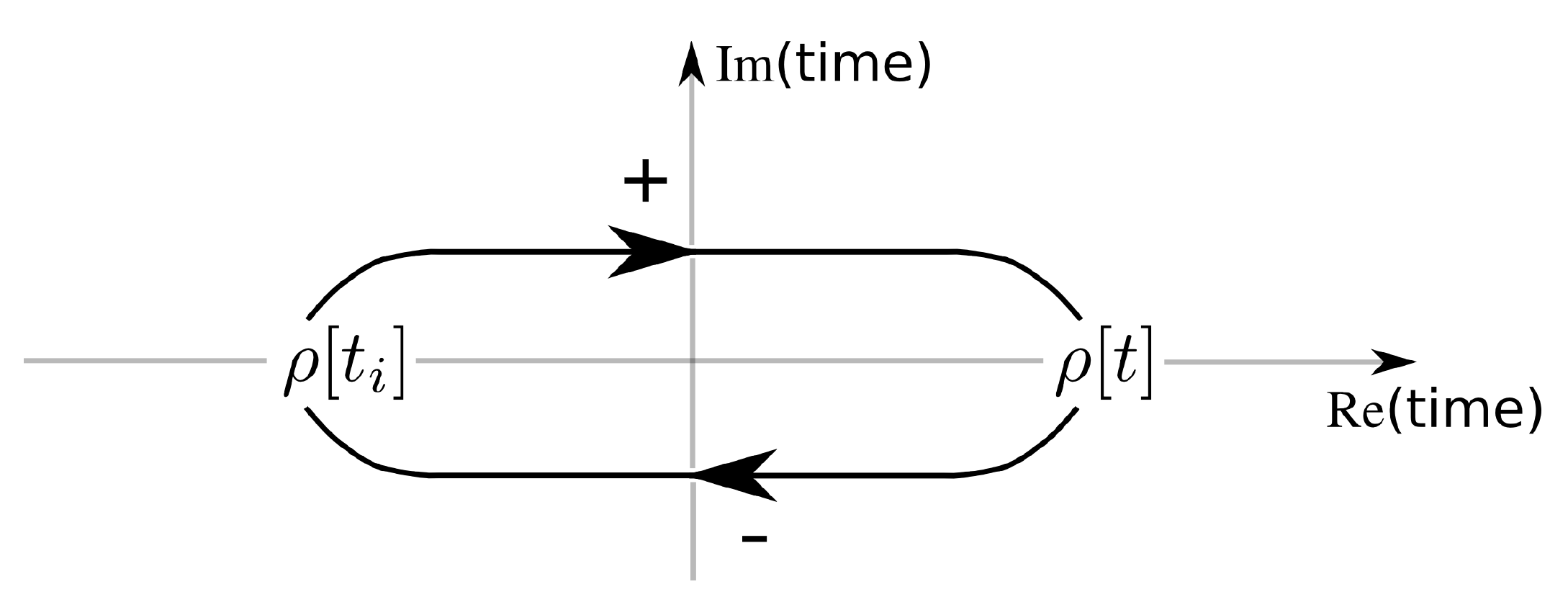}
		\caption{Schematic representation of the time evolution of the system density matrix. Here, $+$ and $-$ label the two branches of the closed-time-path contour.}
		\label{fig:Path}
	\end{center}
\end{figure}

In order to proceed further, we assume that the state of the full system at the initial time $t_i$, i.e.~before the quench, is a product state of the form
\begin{equation}
	\hat{\rho}(t_i) = \hat{\rho}_\phi(t_i) \otimes \hat{\rho}_\chi(t_i)\;.
	\label{eq:rho sep}
\end{equation}
For the setup that we have in mind, this is the assumption that the initial momentum configuration of the ``atom'' subsystem can be prepared as a pure state. The state of the subsystem at time $t$ is then given by
\begin{equation}
	\rho_\phi[\phi_t^{\pm};t] = \int {\rm d}\phi^{\pm}_i\; \mathcal{I}[\phi^{\pm}_t,\phi^{\pm}_i;t,t_i] \rho_\phi[\phi^{\pm}_i;t_i]\;,
	\label{eq:rho IF prop}
\end{equation}
where
\begin{equation}
	\mathcal{I}[\phi^{\pm}_t,\phi^{\pm}_i;t,t_i]\ =\ \int_{\phi_i^{\pm}}^{\phi_t^{\pm}} \mathcal{D}\phi^{\pm}\; {\rm e}^{\frac{i}{\hbar} \widehat{S}_{\rm eff}[\phi;t]}
	\label{eq:IF prop}
\end{equation}
is the influence functional (IF) propagator. The latter arises from inserting complete sets of field and conjugate-momentum eigenstates into Eq.~\eqref{eq:reduceddensfunc} at all intermediate times from the final time $t$, to the initial time $t_i$ \emph{and back again}, giving rise to the closed-time path~\cite{Schwinger:1960qe, Keldysh:1964ud}, pictured in Fig.~\ref{fig:Path}. 

The master equation follows straightforwardly from taking the partial time derivative of Eq.~\eqref{eq:rho IF prop}. If only local interactions are induced within the open subsystem, the master equation takes the form
\begin{equation}
\label{eq:functionalLiouville}
\partial_t\rho_\phi[\phi_t^{\pm};t] \ =\ \frac{i}{\hbar}\,\partial_t\widehat{S}_\text{eff}[\phi;t]\rho_\phi[\phi_t^{\pm};t]\;,
\end{equation}
where the effective action $\widehat{S}_{\rm eff}$ arises from tracing out the $\chi$ field. It takes the general form
\begin{equation}
\label{eq:Seffdef}
\widehat{S}_{\rm eff}[\phi;t]\ =\ \widehat{S}_{\phi}[\phi;t]\:+\:\widehat{S}_{\rm IF}[\phi;t]\;,
\end{equation}
where we use a $\widehat{\ }$ to indicate functionals (and later operators; cf.~Sec.~\ref{sec:TFD}) that depend on both of the doubled field variables $\phi^+$ and $\phi^-$. Contributions to the effective action associated with unitary evolution can be written in terms of the usual action, e.g.,
\begin{equation}
\label{eq:unitaryS}
\widehat{S}_{\phi}[\phi;t]\ =\ \sum_{a\,=\,\pm}a S_{\phi}[\phi^a;t]\ =\ S_{\phi}[\phi^+;t]\:-\:S_{\phi}[\phi^-;t]\;.
\end{equation}
On the other hand, the influence action $\widehat{S}_{\rm IF}[\phi;t]$, which, in general, describes nonunitary dynamics, cannot be constructed in this way. Instead, it will contain terms that mix $+$ and $-$ field variables. 
The partial time derivative of $-\widehat{S}_{\rm eff}$ is nothing other than the Liouvillian $\widehat{H}_{\rm eff}$, such that Eq.~\eqref{eq:functionalLiouville} is just the functional expression of the quantum Liouville equation, including both the Hamiltonian and Lindblad terms. In the case of unitary evolution, we have
\begin{equation}
\label{eq:partialtS}
-\partial_t\widehat{S}\ =\ \sum_{a\,=\,\pm}a\dot{\phi}^a\pi^a\:-\:\widehat{L}\ \equiv\ \widehat{H}\;,
\end{equation}
where $\pi^a=\dot{\phi}^a$ are the canonical conjugate momenta.

The explicit form of $\widehat{S}_{\rm IF}$ depends on the particular interactions and is defined by means of the Feynman-Vernon influence functional~\cite{Calzetta_1998}
\begin{align} \label{eq:Feynman}
&\widehat{\mathcal{F}}[\phi;t]\ =\ \exp\left\{\frac{i}{\hbar}\,\widehat{S}_{\text{IF}}[\phi;t]\right\} \nonumber
\\
&\ = \int {\rm d}\chi^{\pm}_t\,{\rm d}\chi^{\pm}_i\; \delta(\chi_t^+-\chi_t^-)\rho_\chi [\chi^{\pm}_i;t_i] \int^{\chi^{\pm}_t}_{\chi^{\pm}_i} \mathcal{D}\chi^{\pm}\exp \left\{ \frac{i}{\hbar}\Big(\widehat{S}_{\chi}[\chi;t] +\widehat{S}_{\chi,{\rm int}}[\chi;t]+ \widehat{S}_{\rm int}[\phi,\chi;t]\Big) \right\}\;.
\end{align}
The influence action does not, in general, include only local interactions, however, and the functional master equation for the model in Eqs. (\ref{eq:Sphi})--(\ref{eq:Sint}) cannot be written as in Eq. (\ref{eq:functionalLiouville}).

If the system is weakly coupled to the environment, $\widehat{S}_{\rm IF}$ can be obtained perturbatively by expanding the right-hand side of Eq.~\eqref{eq:Feynman}, in our case with respect to the small parameters $\{ \lambda,m/\mathcal{M}, X/\mathcal{M}\}\ll 1$. Expanding to quadratic order in the action, we obtain
\begin{align} \label{eqn:InfluenceActionExpanded}
\widehat{S}_{\text{IF}}[\phi] \ &=\ \sum_{a\,=\,\pm}a\Big[\braket{S_{\rm int}[\phi^a,\chi^a]} + \braket{S_{\chi,{\rm int}}[\chi^a]}\Big] + \frac{i}{2\hbar}\sum_{a,b\,=\,\pm}ab\Big[\braket{S_{\rm int}[\phi^a,\chi^a]S_{\rm int}[\phi^b,\chi^b]}'
\nonumber\\
&\qquad
+ \braket{S_{\chi,{\rm int}}[\chi^a]S_{\chi,{\rm int}}[\chi^b]}'+2\braket{S_{\chi,{\rm int}}[\chi^a]S_{\rm int}[\phi^b,\chi^b]}'\Big]\;,
\end{align}
where
\begin{equation}
\braket{A[\chi^{a}]}\ \equiv\ \int {\rm d}\chi^{\pm}_t\,{\rm d}\chi^{\pm}_i\; \delta(\chi_t^+-\chi_t^-)\rho_\chi [\chi^{\pm}_i;t_i] \int^{\chi^{\pm}_t}_{\chi^{\pm}_i} \mathcal{D}\chi^{\pm}\;A[\chi^{a}]\exp \left\{ \frac{i}{\hbar}\,\widehat{S}_{\chi}[\chi;t] \right\}
\end{equation}
and
\begin{equation}
\braket{A[\chi^{a}]B[\chi^{b}]}'\ \equiv \ \braket{A[\chi^{a}]B[\chi^{b}]}-\braket{A[\chi^{a}]}\braket{B[\chi^{b}]}\;.
\end{equation}
We suppress the time arguments of the contributions to the action when convenient to do so.

In general, it may not be possible to thermally isolate the chameleon fluctuations from the walls of the vacuum chamber. We remark, however, that any such thermal corrections to the dynamics of the screening field are usually assumed to be negligible in atom-interferometry tests. We will comment on this further in Sec.~\ref{sec:discussion}. Nevertheless, in order to keep track of potentially important thermal corrections,
we take the initial state of the environment to be a thermal state (with respect to the fluctuations $\chi$ around the background field $\braket{X}$, taken here to be in equilibrium with the vacuum chamber), in which case
\begin{equation}
\rho_{\chi}[\chi_i^{\pm};t_i]\ =\ \frac{1}{{\rm Tr}\,e^{-\beta \hat{H}_{\chi}}}\braket{\chi^+_i|e^{-\beta \hat{H}_{\chi}}|\chi^-_i}\;,
\end{equation}
where $\beta=1/T$ is the inverse thermodynamic temperature and $\hat{H}_{\chi}$ is the free Hamiltonian of the $\chi$ fluctuations.  We work in units where the Boltzmann constant is unity. The various correlation functions can be evaluated by means of Wick's theorem, and we have (see, e.g., Ref.~\cite{LeBellac})
\begin{subequations}
\label{eq:chiprops}
\begin{flalign}
&\overbracket{\chi^+_x\chi^+_y}\ =\ \braket{{\rm T}[\chi_{x} \chi_{y}]} \ =\  \Delta^{++}_{xy}\ = \ \Delta^{\rm F}_{xy}\nonumber\\&\phantom{\overbracket{\chi^+_x\chi^+_y}\ }\, =\ -\, i\hbar \int_k e^{ik\cdot(x-y)}\bigg[\frac{1}{k^2 + M^2 -i\epsilon}+2\pi i f(|k^0|)\delta(k^2+M^2)\bigg]\;,
\\
&\overbracket{\chi^-_x\chi^-_y}\ =\ \braket{\bar{\rm T}[ \chi_{x}\chi_{y}]}\ =\  \Delta^{--}_{xy}\ =\ \Delta^{\rm D}_{xy}\nonumber\\&\phantom{\overbracket{\chi^-_x\chi^+_y}\ }\, =\ +\, i\hbar \int_k  e^{ik\cdot(x-y)}\bigg[\frac{1}{k^2 + M^2 +i\epsilon}-2\pi i f(|k^0|)\delta(k^2+M^2)\bigg]\;,
\\
&\overbracket{\chi^+_x\chi^-_y}\ =\ \braket{\chi_{y} \chi_{x}}\  =\  \Delta^{+-}_{xy}\ =\ \Delta^<_{xy}\  =\ \hbar\int_k e^{ik\cdot (x-y)} 2\pi{\rm sgn(k^0)}f(k^0)\delta(k^2 + M^2)\;,
\\[0.6em]
&\overbracket{\chi^-_x\chi^+_y}\ =\  \braket{\chi_{x} \chi_{y}}\  =\  \Delta^{-+}_{xy}\ =\ \Delta^>_{xy}\ =\ \Delta^<_{yx}\ =\ (\Delta^<)^*_{xy}\;,
\end{flalign}
\end{subequations}
where $\Delta^{{\rm F}({\rm D})}_{xy}$ is the Feynman (Dyson) propagator, $\Delta^\gtrless$ are the Wightman propagators, and ${\rm T}$ and $\bar{\rm T}$ are the time- and anti-time-ordering operators, respectively, and we have used the shorthand notation
\begin{equation}
\int_k\ \equiv\ \int\!\frac{{\rm d}^4k}{(2\pi)^4}\;.
\end{equation}
The thermal contributions are encoded in the on-shell terms proportional to the Bose-Einstein distribution function
\begin{equation}
f(k^0)\ =\ \frac{1}{e^{\beta k^0}-1}\;.
\end{equation}
Here, ${\rm sgn}(k^0)=\theta(k^0)-\theta(-k^0)$ is the signum function, and the form of the positive-frequency Wightman propagator $\Delta^>_{xy}$ follows from the identity
\begin{equation}
\label{eq:BEiden}
f(-k^0)\ =\ -\big[1+f(k^0)\big]\;.
\end{equation}

The various terms in $\widehat{S}_{\rm IF}$, Eq.~\eqref{eqn:InfluenceActionExpanded}, can then be expressed as follows:
\begin{subequations}
\label{eq:Scontracts}
\begin{align}
&\braket{S_{\rm int}[\phi^a,\chi^a]}\ =\ -\,\frac{m^2}{\mathcal{M}^2}\int_x \big(\phi^a_{x}\big)^2 \Delta^{\rm F}_{xx}\;,\\
&\braket{S_{\rm int}[\phi^a,\chi^a]S_{\rm int}[\phi^b,\chi^b]}'\ =\ \braket{S_{\rm int}[\phi^a,\chi^a]S_{\rm int}[\phi^b,\chi^b]}\ = \ \frac{m^4}{\mathcal{M}^2}\int_{xy}  \big(\phi^a_{x}\big)^2\big(\phi^b_{y}\big)^2 \Delta^{ab}_{xy}\;,\\
&\braket{S_{\chi,{\rm int}}[\chi^a]}\ =\ -\,\frac{\lambda}{4!}\int_{x} \Big[3\big(\Delta^{\rm F}_{xx}\big)^2 +\braket{X}^4\Big]\;, \\
&\braket{S_{\chi,{\rm int}}[\chi^a]S_{\chi,{\rm int}}[\chi^b]}\ =\ \frac{\lambda^2}{(4!)^2}\int_{xy} \Big[24\big(\Delta^{ab}_{xy}\big)^4 + 72\big(\Delta^{\rm F}_{xx}\big)^2\big(\Delta^{ab}_{xy}\big)^2   + 96\braket{X}^2\big(\Delta^{ab}_{xy}\big)^3  &
\nonumber\\
&\qquad
 + 144\braket{X}^2\big(\Delta^{\rm F}_{xx}\big)^2\Delta^{ab}_{xy}
 + 9 \big(\Delta^{\rm F}_{xx}\big)^4 
+ 6\braket{X}^4  \big(\Delta^{\rm F}_{xx}\big)^2+ \braket{X}^8\Big]\;,\\
&\braket{S_{\chi,{\rm int}}[\chi^a]S_{\chi,{\rm int}}[\chi^b]}'\ =\ \frac{\lambda^2}{(4!)^2}\int_{xy} \Big[24\big(\Delta^{ab}_{xy}\big)^4 + 72\big(\Delta^{\rm F}_{xx}\big)^2\big(\Delta^{ab}_{xy}\big)^2 + 96\braket{X}^2\big(\Delta^{ab}_{xy}\big)^3  &
\nonumber\\
&\qquad
 + 144\braket{X}^2\big(\Delta^{\rm F}_{xx}\big)^2\Delta^{ab}_{xy}
 \Big]\;,\\
&\braket{S_{\chi,{\rm int}}[\chi^a] S_{\rm int}[\phi^b,\chi^b]}'\, =\, \braket{S_{\chi,{\rm int}}[\chi^a] S_{\rm int}[\phi^b,\chi^b]}\, =\, \frac{\lambda \braket{X} m^2}{2\mathcal{M}}\int_{xy} 
 \Delta^{\rm F}_{xx}\Delta^{ab}_{xy}\big(\phi^b_y\big)^2\;,
\end{align}
\end{subequations}
where we have restored the factors arising from $V^{\rm eff}(\braket{X})$, omitted in Eq.~\eqref{eq:Schiint}, in order to illustrate that these do not contribute to $\widehat{S}_{\rm IF}$. Hereafter, we leave it implicit that all time integrals run over the domain $[0,t]$. We note that we work in a regime where $\alpha_2^2\ll \alpha_2,\alpha_1^2$, i.e.~where $\braket{X}/\mathcal{M},m/\mathcal{M}\ll 1$. Putting everything together, we have
\begin{align}\label{eqn:InfluenceAction}
\widehat{S}_{\text{IF}}[\phi;t]  \ &= \ -\,\frac{m^2}{\mathcal{M}^2}\sum_{a\,=\,\pm}\int_{x} a\big(\phi^a_x\big)^2\Delta^{\rm F}_{xx} +\frac{i}{2\hbar}\int_{xy} \sum\limits_{a,\,b\,=\,\pm}ab\,\Bigg\{ \frac{m^4}{\mathcal{M}^2}\big(\phi^a_x\big)^2\big(\phi^b_y\big)^2\Delta^{ab}_{xy}   
\nonumber
\\
&\qquad
+\frac{\lambda^2}{24}\Big[\big(\Delta^{ab}_{xy}\big)^4 + 3 \big(\Delta^{\rm F}_{xx}\big)^2\big(\Delta^{ab}_{xy}\big)^2 + 4\braket{X}^2 \big(\Delta^{ab}_{xy}\big)^3+ 6\braket{X}^2\big(\Delta^{\rm F}_{xx}\big)^2\Delta^{ab}_{xy}\Big]\nonumber\\&\qquad
+\frac{\lambda \braket{X} m^2}{\mathcal{M}}\,\Delta^{\rm F}_{xx} \Delta^{ab}_{xy}\big(\phi^b_y\big)^2
\Bigg\}\;.
\end{align}
We hereafter work in natural units, setting $\hbar=1$.

The terms in square brackets in the second line of Eq.~\eqref{eqn:InfluenceAction} do not involve any fields $\phi$. Since it involves only $\chi$ propagators, the sum over all $a$ and $b$ is equivalent to a sum over all underlinings in the sense of the largest time equation~\cite{tHooft:1973wag, Kobes:1985kc}, yielding a vanishing contribution; that is to say
\begin{equation}
\forall \ n\ \in\ \mathbb{N}_0:\qquad  \sum_{a,\,b\,=\,\pm}a b\big(\Delta^{ab}_{xy}\big)^n\ =\ 0\;.
\end{equation}
 Thus, we have that
\begin{align}\label{eqn:InfluenceAction2}
&\widehat{S}_{\text{IF}}[\phi;t]  \ = \ -\,\frac{m^2}{\mathcal{M}^2}\sum_{a\,=\,\pm}\int_{x} a\big(\phi^a_x\big)^2\Delta^{\rm F}_{xx} +\frac{i}{2\hbar}\int_{xy} \sum\limits_{a,\,b\,=\,\pm}ab\,\Bigg\{ \frac{m^4}{\mathcal{M}^2}\big(\phi^a_x\big)^2\big(\phi^b_y\big)^2\Delta^{ab}_{xy}   
\nonumber
\\
&\quad
+\frac{\lambda \braket{X} m^2}{\mathcal{M}}\,\Delta^{\rm F}_{xx}\Delta^{ab}_{xy}\big(\phi^b_y\big)^2
\Bigg\}\;.
\end{align}


\subsection{Operator-based approach}
\label{sec:TFD}

Next, we wish to extract from the functional expression of the quantum Liouville equation, as obtained from the partial time derivative of Eq. (\ref{eq:rho IF prop}) via Eqs. (\ref{eq:IF prop}) and (\ref{eq:Feynman}), matrix elements of the reduced density operator in the bases of momentum eigenstates. In analogy with scattering-matrix calculations, we therefore need a method of LSZ reduction. In order to guide such a procedure in the non-equilibrium setting, it is helpful to consider the corresponding canonical, operator-based formulation of non-equilibrium field theory.
This is known as thermo field dynamics (TFD)~\cite{Takahasi:1974zn, Arimitsu:1985ez, Arimitsu:1985xm} (see also Ref.~\cite{Khanna}), and the doubling of field degrees of freedom needed to describe relativistic quantum statistical systems requires us to construct this canonical formalism over a doubled Hilbert space $\widehat{\mathscr{H}}\equiv \mathscr{H}^+\otimes\mathscr{H}^-$. The usual scalar field operator $\hat{\phi}$ is then embedded by defining the plus- and minus-type field operators
\begin{equation}
\hat{\phi}^+(x)\ \equiv\ \hat{\phi}(x)\:\otimes\:\hat{\mathbb{I}}\;,\qquad
\hat{\phi}^-(x)\ \equiv\ \hat{\mathbb{I}}\:\otimes\:\hat{\phi}^{\mathcal{T}}(x)\;,
\end{equation}
with analogous expressions for the embeddings of the usual scalar creation and annihilation operators. Here, the $\mathcal{T}$ indicates time reversal. The interaction-picture field operators can then be written in the usual plane-wave decompositions
\begin{equation}
\hat{\phi}^{\pm}(x)\ =\ \int{\rm d}\Pi_{\mathbf{k}}\;\Big[\hat{a}^{\pm}_{\mathbf{k}}\,e^{\mp i E_{\mathbf{k}}t\pm i\mathbf{k}\cdot\mathbf{x}}\:+\:\hat{a}^{\pm\dag}_{\mathbf{k}}\,e^{\pm i E_{\mathbf{k}}t\mp i\mathbf{k}\cdot \mathbf{x}}\Big]\;,
\end{equation}
where we use the notation
\begin{equation}
\int{\rm d}\Pi_{\mathbf{k}}\ \equiv\ \int_{\mathbf{k}}\frac{1}{2E_{\mathbf{k}}}\;,\qquad \int_{\mathbf{k}}\ \equiv\ \int\!\frac{{\rm d}^3\mathbf{k}}{(2\pi)^3}\;,
\end{equation}
for the Lorentz-invariant phase-space integrals. When no time arguments are provided, it is assumed that the operators and states are evaluated at $t=0$. Notice that the minus-type field is built from the time-reversed field operator, reflecting the anti-time-ordering of the negative branch of the closed-time path in the path-integral formulation. We also note that, by virtue of the Kronecker product structure, the plus- and minus-type operators commute with one another.

The plus- and minus-type creation and annihilation operators act on states in the corresponding Hilbert spaces (and Fock spaces), that is, by acting on the doubled vacuum state $\kket{0}\equiv\ket{0}\otimes\ket{0}$, we have
\begin{equation}
\label{eq:creation}
\hat{a}^{+\dag}_{\mathbf{k}}\kket{0}\ =\ \ket{\mathbf{k}}\:\otimes\:\ket{0}\ \equiv\ \kket{\mathbf{k}_+}\;,\qquad
\hat{a}^{-\dag}_{\mathbf{k}}\kket{0}\ =\ \ket{0}\:\otimes\:\ket{\mathbf{k}}\ \equiv\ \kket{\mathbf{k}_-}\;,
\end{equation}
and
\begin{equation}
\label{eq:annihilation}
\hat{a}^{+}_{\mathbf{k}}\kket{\mathbf{p}_+,\mathbf{p}_-}\ =\ (2\pi)^32E_{\mathbf{k}}\delta^{(3)}(\mathbf{p}-\mathbf{k})\kket{\mathbf{p}_-}\;,\qquad
\hat{a}^{-}_{\mathbf{k}}\kket{\mathbf{p}_+,\mathbf{p}_-}\ =\ (2\pi)^32E_{\mathbf{k}}\delta^{(3)}(\mathbf{p}-\mathbf{k})\kket{\mathbf{p}_+}\;,
\end{equation}
where $\kket{\mathbf{p}_+,\mathbf{p}_-}\equiv\ket{\mathbf{p}}\:\otimes\:\ket{\mathbf{p}}$ and so on.

The density operator of an isolated system can be embedded as
\begin{equation}
\hat{\rho}^+(t)\ \equiv\ \hat{\rho}(t)\:\otimes\:\hat{\mathbb{I}}\;,
\end{equation}
where $\hat{\mathbb{I}}$ is the unit operator. Its trace can then be expressed in the following form:
\begin{equation}
{\rm tr}\,\hat{\rho}(t)\ =\ \bbrakket{1|\hat{\rho}^+(t)|1}\;,
\end{equation}
where the state (see Ref.~\cite{Arimitsu:1985ez})
\begin{equation}
\kket{1}\ \equiv \ \kket{0}\:+\:\int{\rm d}\Pi_{\mathbf{p}_1}\;\kket{\mathbf{p}_{1+},\mathbf{p}_{1-}}\:+\:\frac{1}{2!}\int{\rm d}\Pi_{\mathbf{p}_1}\,{\rm d}\Pi_{\mathbf{p}_2}\;\kket{\mathbf{p}_{1+},\mathbf{p}_{2+},\mathbf{p}_{1-},\mathbf{p}_{2-}}\:+\:\dots\;.
\end{equation}
The trace of an operator $\hat{O}(t)$ can be written
\begin{equation}
{\rm tr}\,\hat{O}(t)\hat{\rho}(t)\ =\ \bbrakket{1|\hat{O}^+(t)\hat{\rho}^+(t)|1}\;.
\end{equation}
In this way, one is able to recast the (unitary) quantum Liouville equation (in the Schr\"{o}dinger picture)
\begin{equation}
\partial_t\hat{\rho}(t)\ =\ -\,i\big[\hat{H},\hat{\rho}(t)\big]
\end{equation}
in the Schr\"{o}dinger-like form
\begin{equation}
\partial_t\hat{\rho}^+(t)\kket{1}\ =\ -\,i\widehat{H}\hat{\rho}^+(t)\kket{1}\;,
\end{equation}
where
\begin{equation}
\widehat{H}\ \equiv\ \hat{H}\:\otimes\:\hat{\mathbb{I}}\:-\:\hat{\mathbb{I}}\:\otimes\:\hat{H}
\end{equation}
is the Liouvillian operator.

Moving to the interaction picture, we take a density operator of the form
\begin{equation}
\hat{\rho}(t)\ =\ \int{\rm d}\Pi_{\mathbf{k}}\,{\rm d}\Pi_{\mathbf{k}'}\;\rho(\mathbf{k},\mathbf{k}';t)\ket{\mathbf{k};t}\bra{\mathbf{k}';t}\;.
\end{equation}
We recall that state and basis vectors evolve respectively with the interaction and free parts of the Hamiltonian in the interaction picture. We assume that the single-particle term dominates in the low-energy, nonrelativistic limit for the matter scalar and set the occupancy of all multiparticle states to zero. We are therefore interested in the matrix element
\begin{equation}
\braket{\mathbf{p};t|\hat{\rho}(t)|\mathbf{p}';t}\ =\ \rho(\mathbf{p},\mathbf{p}';t)\;.
\end{equation}
Notice that all of the basis states and operators are evaluated at equal times, and the matrix element $\rho(\mathbf{p},\mathbf{p}';t)$ is therefore picture independent (see Refs.~\cite{Millington:2012pf, Millington:2013isa}).
In the TFD language, this can be written as
\begin{equation}
{\rm tr}\ket{\mathbf{p}';t}\!\bra{\mathbf{p};t}\hat{\rho}(t)\ =\ \bbrakket{1(t)|(\ket{\mathbf{p}';t}\!\bra{\mathbf{p};t}\otimes\hat{\mathbb{I}})(\hat{\rho}(t)\otimes\hat{\mathbb{I}})|1(t)}\;,
\end{equation}
and we draw attention to the time dependence of the state $\kket{1(t)}$. Using the fact that
\begin{equation}
\bbra{1(t)}(\ket{\mathbf{p}';t}\!\bra{\mathbf{p};t}\otimes\hat{\mathbb{I}})\ =\ \bbra{\mathbf{p}_+,\mathbf{p}_-';t}
\end{equation}
and
\begin{equation}
(\hat{\rho}(t)\otimes\hat{\mathbb{I}})\kket{1(t)}\ =\ \int{\rm d}\Pi_{\mathbf{k}}\,{\rm d}\Pi_{\mathbf{k}'}\;\rho(\mathbf{k},\mathbf{k}';t)\kket{\mathbf{k}_+,\mathbf{k}_-';t}\;,
\end{equation}
it follows that
\begin{equation}
\partial_t\rho(\mathbf{p},\mathbf{p}';t)\ =\ -\,i\int{\rm d}\Pi_{\mathbf{k}}\,{\rm d}\Pi_{\mathbf{k}'}\;\rho(\mathbf{k},\mathbf{k}';t)\bbrakket{\mathbf{p}_+,\mathbf{p}_-';t|\widehat{H}(t)|\mathbf{k}_+,\mathbf{k}_-';t}\;.
\end{equation}
This can be rewritten as
\begin{equation}
\label{eq:hhatmiddle}
\partial_t \rho(\mathbf{p},\mathbf{p}';t)\ =\ -\,i\int{\rm d}\Pi_{\mathbf{k}}\,{\rm d}\Pi_{\mathbf{k}'}\;\rho(\mathbf{k},\mathbf{k}';t)\bbrakket{0|\hat{a}^+_{\mathbf{p}}(t)\hat{a}^-_{\mathbf{p}'}(t)\widehat{H}(t)\hat{a}_{\mathbf{k}}^{+\dag}(t)\hat{a}_{\mathbf{k}'}^{-\dag}(t)|0}\;.
\end{equation}
In our case, $\widehat{H}\to \widehat{H}_{\rm eff}=  -\,\partial_t\widehat{S}_{\rm eff}$, cf.~Eq.~\eqref{eq:partialtS} is the effective (and non-Hermitian) Liouvillian that comes from tracing out the chameleon degrees of freedom.
Allowing for the fact that $\widehat{H}_{\rm eff}$ is a nonlocal, but time-ordered operator, and after accounting for the free-phase evolution of the rightmost creation operators, the expectation value on the right-hand side of Eq.~\eqref{eq:hhatmiddle} can be time ordered as
\begin{equation}
\label{eq:rhodotFock}
\partial_t\rho(\mathbf{p},\mathbf{p}';t)\ =\ -\,i\int{\rm d}\Pi_{\mathbf{k}}\,{\rm d}\Pi_{\mathbf{k}'}\;e^{i(E_{\mathbf{k}}-E_{\mathbf{k}'})t}\rho(\mathbf{k},\mathbf{k}';t)\bbrakket{0|{\rm T}[\hat{a}^+_{\mathbf{p}}(t)\hat{a}^-_{\mathbf{p}'}(t)\widehat{H}_{\rm eff}(t)\hat{a}_{\mathbf{k}}^{+\dag}(0)\hat{a}_{\mathbf{k}'}^{-\dag}(0)]|0}\;.
\end{equation}

Continuing, we can recast the expression in terms of field operators by using
\begin{align}
\hat{a}^+_{\mathbf{p}}(t)\ =\ +\,i\int_{\mathbf{x}}\;e^{-i\mathbf{p}\cdot\mathbf{x}}\partial_{t,E_{\mathbf{p}}}\hat{\phi}^+(t,\mathbf{x})\;,\qquad 
\hat{a}^{+\dag}_{\mathbf{p}}(t)\ =\ -\,i\int_{\mathbf{x}}\;e^{+i\mathbf{p}\cdot\mathbf{x}}\partial_{t,E_{\mathbf{p}}}^*\hat{\phi}^+(t,\mathbf{x})\;,
\end{align}
where
\begin{equation}
\partial_{t,E_{\mathbf{p}}}\ \equiv\ \overset{\rightarrow}{\partial}_t\:-\:iE_{\mathbf{p}}\;,
\end{equation}
along with
\begin{align}
\hat{a}^-_{\mathbf{p}}(t)\ =\ -\,i\int_{\mathbf{x}}\;e^{+i\mathbf{p}\cdot\mathbf{x}}\partial_{t,E_{\mathbf{p}}}^*\hat{\phi}^-(x)\;,\qquad
\hat{a}^{-\dag}_{\mathbf{p}}(t)\ =\ +\,i\int_{\mathbf{x}}\;e^{-i\mathbf{p}\cdot\mathbf{x}}\partial_{t,E_{\mathbf{p}}}\hat{\phi}^{-}(x)\;.
\end{align}
Specifically, we have
\begin{align}
\label{eq:final_op_reduc}
\partial_t\rho(\mathbf{p},\mathbf{p}';t)\ &=\ -\,i\lim_{\substack{x^{0(\prime)}\,\to\, t^{+}\\y^{0(\prime)}\,\to\, 0^-}}\int{\rm d}\Pi_{\mathbf{k}}\,{\rm d}\Pi_{\mathbf{k}'}\;e^{i(E_{\mathbf{k}}-E_{\mathbf{k}'})t}\rho(\mathbf{k},\mathbf{k}';t)\int_{\mathbf{x}\mathbf{x}'\mathbf{y}\mathbf{y'}}e^{-i(\mathbf{p}\cdot\mathbf{x}-\mathbf{p}'\cdot\mathbf{x}')+i(\mathbf{k}\cdot\mathbf{y}-\mathbf{k}'\cdot\mathbf{y}')}\nonumber\\&\qquad\times\:\partial_{x^0,E_{\mathbf{p}}}\partial_{x^{0\prime},E_{\mathbf{p}'}}^*\partial_{y^0,E_{\mathbf{k}}}^*\partial_{y^{0\prime},E_{\mathbf{k}'}}\bbrakket{0|{\rm T}[\hat{\phi}^+(x)\hat{\phi}^-(x')\widehat{H}_{\rm eff}(t)\hat{\phi}^{+}(y)\hat{\phi}^{-}(y')]|0}\;,
\end{align}
where $x^{0(\prime)}$ approaches $t$ from above and $y^{0(\prime)}$ approaches $0$ from below to ensure that the time ordering of the operators is equivalent to the original ordering in Eq.~\eqref{eq:rhodotFock}. Notice that the role of the differential operators is to chop off external propagators and replace them with plane-wave factors. The procedure outlined here for projecting into the single-particle subspace therefore amounts to an LSZ-like reduction~\cite{Lehmann:1954rq} of the four-point function
\begin{equation}
\label{eq:4point}
\bbrakket{0|{\rm T}[\hat{\phi}^+(x)\hat{\phi}^-(x')\widehat{H}_{\rm eff}(t)\hat{\phi}^{+}(y)\hat{\phi}^{-}(y')]|0}\;.
\end{equation}
We remark that this procedure readily generalizes to density matrices that include momentum elements of different multiplicity. For instance, projecting onto the \emph{up-to}-two-particle subspace, one could obtain additional elements $\rho(\mathbf{p};t)$, $\rho(\mathbf{p},\mathbf{p}',\mathbf{p}'';t)$ and $\rho(\mathbf{p},\mathbf{p}',\mathbf{p}'',\mathbf{p}''';t)$, which would be obtained respectively from the reduction of two-, six- and eight-point functions analogous to Eq.~\eqref{eq:4point}.

In the path-integral language, Eq.~\eqref{eq:final_op_reduc} can be recast as
\begin{align}
\label{eq:final_path_reduc}
&\partial_t\rho(\mathbf{p},\mathbf{p}';t)\ =\ -\,i\lim_{\substack{x^{0(\prime)}\,\to\, t^{+}\\y^{0(\prime)}\,\to\, 0^-}}\int{\rm d}\Pi_{\mathbf{k}}\,{\rm d}\Pi_{\mathbf{k}'}\;e^{i(E_{\mathbf{k}}-E_{\mathbf{k}'})t}\rho(\mathbf{k},\mathbf{k}';t)\int_{\mathbf{x}\mathbf{x}'\mathbf{y}\mathbf{y'}}e^{-i(\mathbf{p}\cdot\mathbf{x}-\mathbf{p}'\cdot\mathbf{x}')+i(\mathbf{k}\cdot\mathbf{y}-\mathbf{k}'\cdot\mathbf{y}')}\nonumber\\&\qquad\times\:\partial_{x^0,E_{\mathbf{p}}}\partial_{x^{0\prime},E_{\mathbf{p}'}}^*\partial_{y^0,E_{\mathbf{k}}}^*\partial_{y^{0\prime},E_{\mathbf{k}'}}\int\mathcal{D}\phi^{\pm}\;e^{i\widehat{S}_{\phi}[\phi]}\phi^+(x)\phi^-(x')\widehat{H}_{\rm eff}(\phi^{\pm})\phi^{+}(y)\phi^{-}(y')\;,
\end{align}
which can be expanded in terms of the $2\times2$ matrix propagator. The latter is of the diagonal form
\begin{equation}
D_{xy}^{ab}\ =\ \begin{bmatrix} D^{\rm F}_{xy} & 0 \\ 0 & D^{\rm D}_{xy}\end{bmatrix}\;,
\end{equation}
since, by virtue of the fact that Eq.~\eqref{eq:final_op_reduc} is a vacuum expectation value, there can be no $+-$ contractions, and the off-diagonal elements of this $2\times2$ matrix propagator are zero. We emphasize that, while $\bbrakket{1|\hat{\phi}^{+(-)}_x\hat{\phi}^{-(+)}_y|0}=D^{<(>)}_{xy}$ (as arises at zero temperature when we do not restrict to the single-particle subspace), $\bbrakket{0|\hat{\phi}^{+(-)}_x\hat{\phi}^{-(+)}_y|0}=0$.

The Feynman and Dyson propagators of the $\phi$ field are
\begin{subequations}
\begin{flalign}
&\bbrakket{0|{\rm T}[\hat{\phi}^+_{x} \hat{\phi}^+_{y}]|0} \ =\ D^{++}_{xy}\ = \ D^{\rm F}_{xy} \ =\ -\, i\hbar \int_k\frac{e^{ik\cdot(x-y)}}{k^2 + m^2 -i\epsilon}\;,
\\
&\bbrakket{0|{\rm T}[ \hat{\phi}^-_{x} \hat{\phi}^-_{y}]|0}\ =\ D^{--}_{xy}\ =\ D^{\rm D}_{xy}\ =\ +\, i\hbar \int_k\frac{e^{ik \cdot(x-y)}}{k^2 + m^2 +i\epsilon}\;,
\end{flalign}
\end{subequations}
Since we have restricted to the single-particle subspace, we necessarily assume that the $\phi$ field remains at zero temperature and therefore out of equilibrium with the environment formed by the conformally coupled scalar $\chi$ [cf.~Eq.~\eqref{eq:chiprops}].


\subsection{Quantum master equation}
\label{sec:qme}

\begin{figure}

\vspace{1em}

\centering

\subfloat[][]{\includegraphics[scale=0.2]{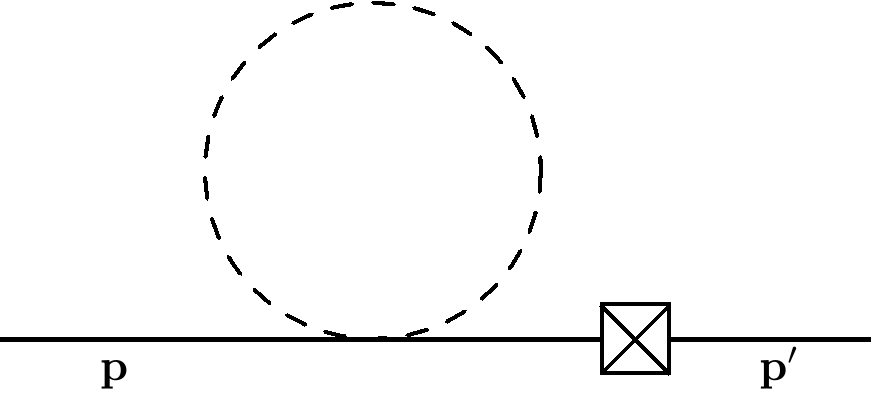}}
\qquad
\subfloat[][]{\includegraphics[scale=0.2]{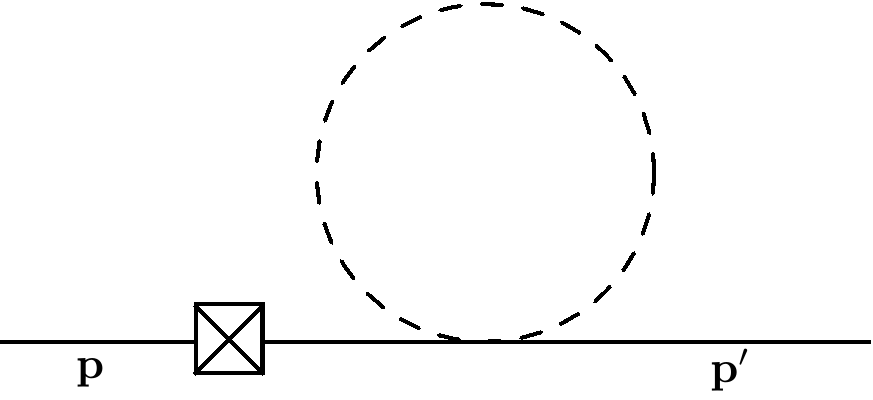}}

\subfloat[][]{\includegraphics[scale=0.2]{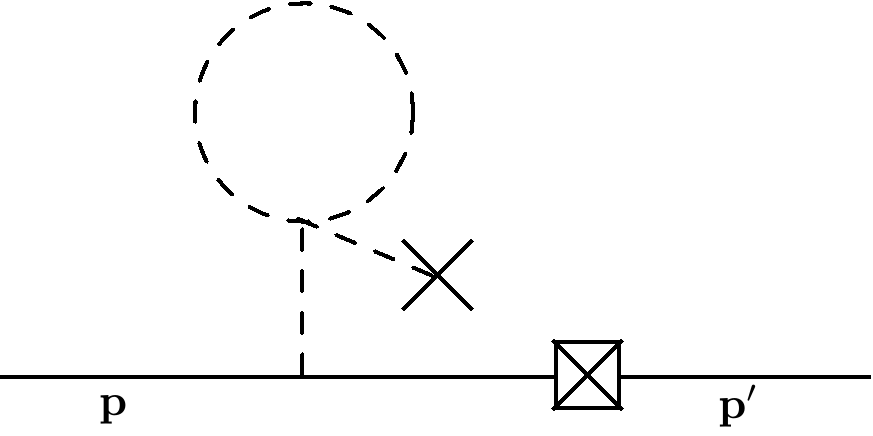}}
\qquad
\subfloat[][]{\includegraphics[scale=0.2]{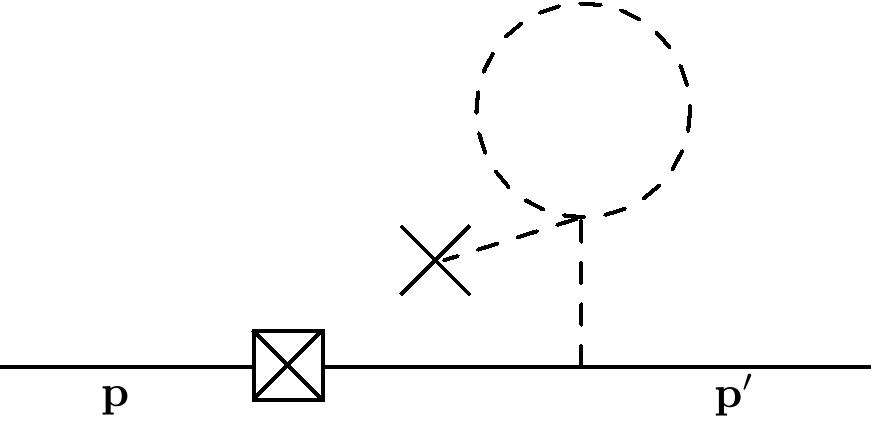}}

\subfloat[][]{\includegraphics[scale=0.2]{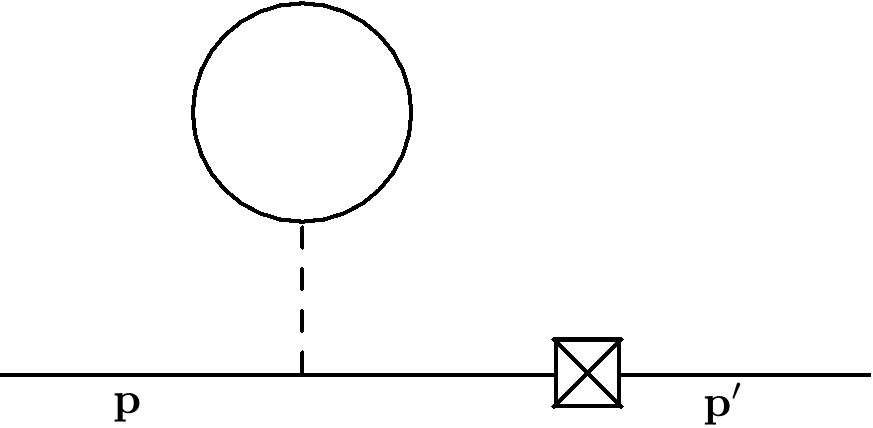}}
\qquad
\subfloat[][]{\includegraphics[scale=0.2]{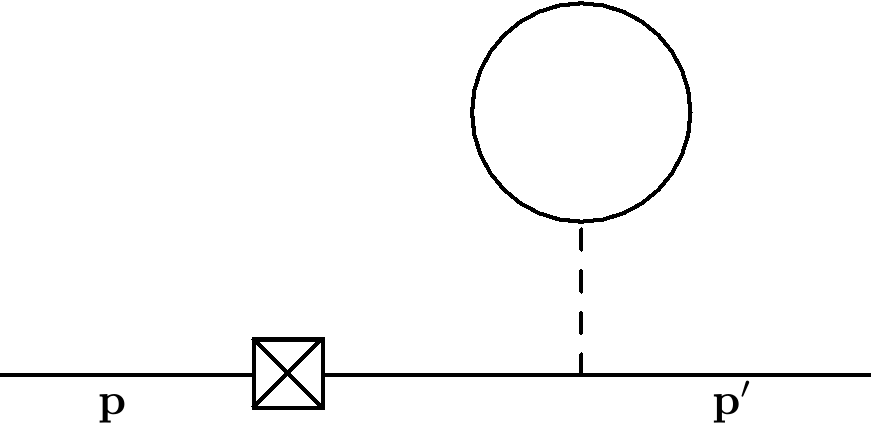}}

\subfloat[][]{\includegraphics[scale=0.2]{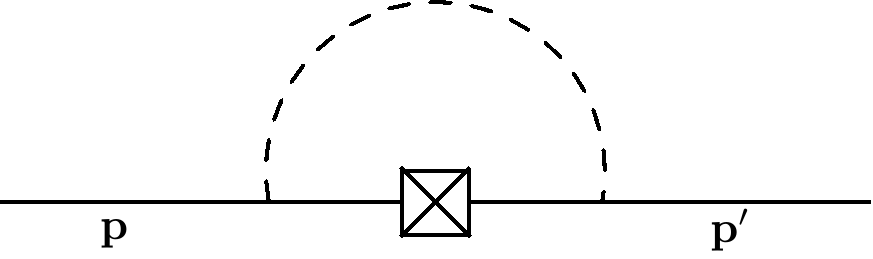}}

\subfloat[][]{\includegraphics[scale=0.2]{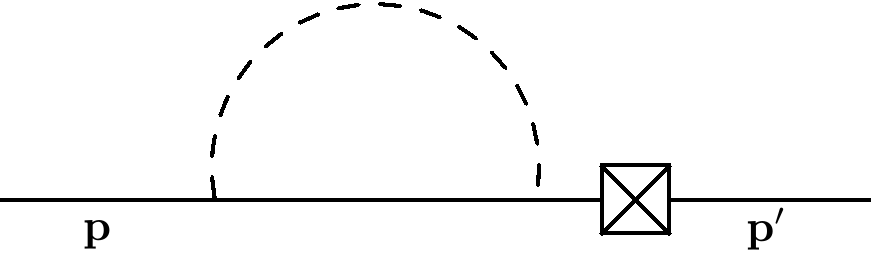}}
\qquad
\subfloat[][]{\includegraphics[scale=0.2]{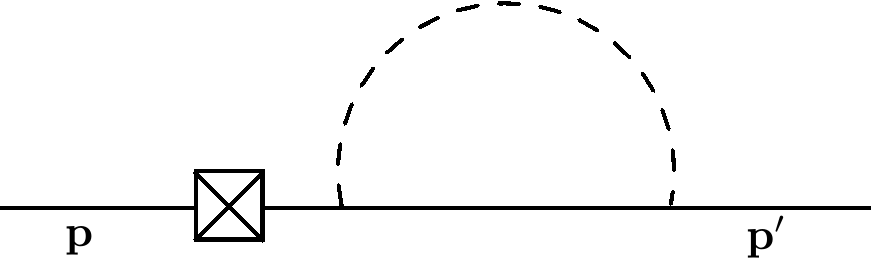}}

\subfloat[][]{\includegraphics[scale=0.175]{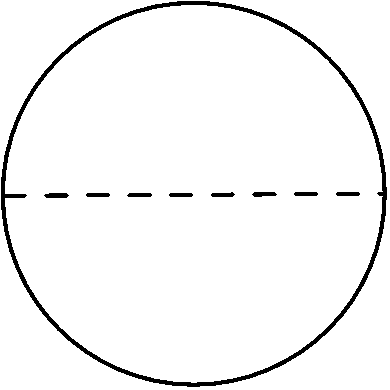}}

\caption{\label{fig:feyns}Diagrammatic representation of the various terms contributing to the right-hand side of the quantum master equation~\eqref{eq:master}: (a)-(d) $\chi$ tadpoles [line 2]; (e) and (f) $\phi$ tadpoles [line 2]; (g)-(i) $\chi$-$\phi$ bubbles [lines 3-5]; and (j) the disconnected vacuum diagram [lines 6 and 7]. Solid lines represent $\phi$ propagators, dashed lines represent $\chi$ propagators, crossed boxes indicate insertions of the matrix element of the density operator $\hat{\rho}$, and crosses indicate insertions of the background field $\braket{X}$.}

\end{figure}

After projecting into the single-particle subspace, as described in Sec.~\ref{sec:TFD}, we arrive at the following quantum master equation:
\begin{align}
\label{eq:startmaster}
&\partial_t\rho(\mathbf{p},\mathbf{p}';t)\ =\ i\lim_{\substack{x^{0(\prime)}\,\to\, t^{+}\\y^{0(\prime)}\,\to\, 0^-}}\int{\rm d}\Pi^{\phi}_{\mathbf{k}}\,{\rm d}\Pi^{\phi}_{\mathbf{k}'}\;e^{i(E^{\phi}_{\mathbf{k}}-E^{\phi}_{\mathbf{k}'})t}\rho(\mathbf{k},\mathbf{k}';t)\int_{\mathbf{x}\mathbf{x}'\mathbf{y}\mathbf{y'}}e^{-i(\mathbf{p}\cdot\mathbf{x}-\mathbf{p}'\cdot\mathbf{x}')+i(\mathbf{k}\cdot\mathbf{y}-\mathbf{k}'\cdot\mathbf{y}')}\nonumber\\&\qquad\times\:\partial_{x^0,E^{\phi}_{\mathbf{p}}}\partial_{x^{0\prime},E^{\phi}_{\mathbf{p}'}}^*\partial_{y^0,E^{\phi}_{\mathbf{k}}}^*\partial_{y^{0\prime},E^{\phi}_{\mathbf{k}'}}\int\mathcal{D}\phi^{\pm}\;e^{i\widehat{S}_{\phi}[\phi]}\phi^+(x)\phi^-(x')\Big(\partial_t\widehat{S}_{\rm eff}[\phi;t]\Big)\phi^{+}(y)\phi^{-}(y')\;,
\end{align}
which we have expressed in terms of the partial time derivative of $\widehat{S}_{\rm eff}$; cf.~Eq.~\eqref{eq:partialtS}. Substituting for the explicit form of the effective action from Eqs.~\eqref{eq:Seffdef}, \eqref{eq:unitaryS}, and \eqref{eqn:InfluenceAction2} in Sec.~\ref{sec:Influence functional}, and performing the remaining Wick contractions, we arrive at
\begin{align}
\label{eq:master}
\partial_t\rho(\mathbf{p},\mathbf{p}';t) \ & =\ -\:i\big(E^{\phi}_{\mathbf{p}}-E^{\phi}_{\mathbf{p}'})\rho(\mathbf{p},\mathbf{p}';t)
\:-\:
\frac{im^2}{\mathcal{M}}\left(\frac{1}{E^{\phi}_{\mathbf{p}}}-\frac{1}{E^{\phi}_{\mathbf{p}'}}\right)\rho(\mathbf{p},\mathbf{p}';t) 
\nonumber
\\
&\quad\quad
\times \bigg\{
\frac{\Delta^{\rm F}_{xx}}{\mathcal{M}}
+
\bigg[\frac{m^2}{\mathcal{M}}D^{\rm F}_{xx}+
\frac{\lambda}{2}\braket{X}\Delta^{\rm F}_{xx}\bigg]\int_{x^0} \frac{\sin[M(x^0-t)]}{M}\bigg\}
\nonumber
\\
&\phantom{=\ }
-
\frac{4m^4}{\mathcal{M}^2}\int_{x^0}\int_{\mathbf{k}}\Bigg\{
\Bigg[\rho(\mathbf{p},\mathbf{p}';t)\,\frac{\cos\big[E^{\phi}_{\mathbf{p}}(t-x^0)\big]\exp\!\big[-iE^{\phi}_{\mathbf{p}-\mathbf{k}}(t-x^0)\big]}{E^{\phi}_{\mathbf{p}}2E^{\chi}_{\mathbf{k}}2E^{\phi}_{\mathbf{p}-\mathbf{k}}}\nonumber\\&\qquad-\rho(\mathbf{p}-\mathbf{k},\mathbf{p}'-\mathbf{k};t) 
\,\frac{\exp\!\big[i\big(E_{\mathbf{p}-\mathbf{k}}^{\phi}-E_{\mathbf{p}}^{\phi}\big)( t-x^0)\big]}{2E^{\chi}_{\mathbf{k}}2E^{\phi}_{\mathbf{p}-\mathbf{k}} 2E^{\phi}_{\mathbf{p}'-\mathbf{k}} }\Bigg]\nonumber\\&\qquad\times\Big[\exp\!\big[-iE_{\mathbf{k}}^{\chi}(t-x^0)\big]+2\cos\big[E_{\mathbf{k}}^{\chi}(t-x^0)\big]f\big(E_{\mathbf{k}}^{\chi}\big)\Big]\: +\:\big(\mathbf{p}\longleftrightarrow \mathbf{p}'\big)^*\Bigg\}\nonumber\\&\qquad -\:\frac{4m^4}{\mathcal{M}^2}\,\rho(\mathbf{p},\mathbf{p}';t)\sum_{s\,=\,\pm}\int_x\int_{\mathbf{k}_1\mathbf{k}_2}\frac{\cos\big[\big(E_{\mathbf{k}_1}^{\phi}+E_{\mathbf{k}_1-\mathbf{k}_2}^{\phi}+sE_{\mathbf{k}_2}^{\chi}\big)(x^0-t)\big]}{2E_{\mathbf{k}_1}^{\phi}2E_{\mathbf{k}_1-\mathbf{k}_2}^{\phi}2E_{\mathbf{k}_2}^{\chi}}\nonumber\\&\qquad\times s\big[1+f\big(sE_{\mathbf{k}_2}^{\chi}\big)\big]\;.
\end{align}
The right-hand side of this expression is represented diagrammatically in Fig.~\ref{fig:feyns}. Figures~\ref{fig:feyns}(a) and \ref{fig:feyns}(b) correspond to the $\chi$ tadpole insertion arising from the first term in the second line of Eq.~\eqref{eq:master}, and \ref{fig:feyns}(c) and \ref{fig:feyns}(d) correspond to the $\chi$ (lollipop) tadpole from the third term in the second line.  Figures~\ref{fig:feyns}(e) and \ref{fig:feyns}(f) correspond to the $\phi$ tadpoles, arising from the second term in the second line, and the remaining Figs.~\ref{fig:feyns}(g)-\ref{fig:feyns}(i) are the $\chi$-$\phi$ bubble diagrams, appearing in the third to fifth lines of Eq.~\eqref{eq:master}. The final two lines of Eq.~\eqref{eq:master} contain a contribution from the absorptive part of the disconnected vacuum diagram shown in Fig.~\ref{fig:feyns}(j). The disconnected diagrams could be absorbed order by order in a redefinition of the matrix element $\rho(\mathbf{p},\mathbf{p}';t)\to \rho(\mathbf{p},\mathbf{p}';t)(1+\text{disconnected diagrams})$, taking into account that the time derivative on the left-hand side counts at a finite order in the coupling constants. For our present discussions, however, we leave the vacuum diagram explicit throughout for completeness.

The terms in the third to seventh lines of Eq.~\eqref{eq:master} include decay (i.e.,~$\chi \to \phi \phi$) and production processes (i.e.,~$\phi\phi \to \chi$), which we would not expect to be present for realistic atoms, which are complex and stable configurations of many elementary particle fields rather than a simple scalar one. The decay and production processes arise here, because such processes are permitted in the simple scalar field theory that we have used as a toy proxy for the atom.

After performing the remaining $x^0$ integral, we obtain
 \begin{align}
\label{eq:masterafterint}
\partial_t\rho(\mathbf{p},\mathbf{p}';t) \ & =\ -\:i\big(E^{\phi}_{\mathbf{p}}-E^{\phi}_{\mathbf{p}'})\rho(\mathbf{p},\mathbf{p}';t)
\:-\:\frac{im^2}{\mathcal{M}}\left(\frac{1}{E^{\phi}_{\mathbf{p}}}-\frac{1}{E^{\phi}_{\mathbf{p}'}}\right)\rho(\mathbf{p},\mathbf{p}';t) 
\nonumber
\\
&\quad\quad
\times \bigg\{
\frac{\Delta^{\rm F}_{xx}}{\mathcal{M}}
+
\bigg[\frac{m^2}{\mathcal{M}}D^{\rm F}_{xx}+
\frac{\lambda}{2}\braket{X}\Delta^{\rm F}_{xx}\bigg]\frac{\cos(Mt)-1}{M^2}\bigg\}
\nonumber
\\
&\phantom{=\ }
+i
\frac{4m^4}{\mathcal{M}^2}\sum_{s\,=\,\pm}\int_{\mathbf{k}}\Bigg\{\Bigg[
\rho(\mathbf{p},\mathbf{p}';t)\,\frac{1}{E^{\phi}_{\mathbf{p}}2E^{\chi}_{\mathbf{k}}2E^{\phi}_{\mathbf{p}-\mathbf{k}}}\,\frac{s}{\big(sE_{\mathbf{k}}^{\chi}+E^{\phi}_{\mathbf{p}-\mathbf{k}}\big)^2-\big(E_{\mathbf{p}}^{\phi}\big)^2}\nonumber\\&\qquad\times\Big[\Big(sE_{\mathbf{k}}^{\chi}+E^{\phi}_{\mathbf{p}-\mathbf{k}}\Big)\Big(1-\exp\big[-i\big(s E_{\mathbf{k}}^{\chi}+E_{\mathbf{p}-\mathbf{k}}^{\phi}\big)t\big]\cos\big(E_{\mathbf{p}}^{\phi}t\big)\Big)\nonumber\\&\qquad-iE_{\mathbf{p}}^{\phi}\exp\big[-i\big(s E_{\mathbf{k}}^{\chi}+E_{\mathbf{p}-\mathbf{k}}^{\phi}\big)t\big]\sin\big(E_{\mathbf{p}}^{\phi}t\big)\Big]\big[1+f\big(sE_{\mathbf{k}}^{\chi}\big)\big]
\nonumber
\\
&
 \qquad +\:\rho(\mathbf{p}-\mathbf{k},\mathbf{p}'-\mathbf{k};t) 
\,\frac{1}{2E^{\chi}_{\mathbf{k}}2E^{\phi}_{\mathbf{p}-\mathbf{k}} 2E^{\phi}_{\mathbf{p}'-\mathbf{k}}}\,\frac{s}{sE_{\mathbf{k}}^{\chi}+E_{\mathbf{p}-\mathbf{k}}^{\phi}-E_{\mathbf{p}}^{\phi}}\nonumber\\&\qquad\times\Big(1-\exp\big[i(sE_{\mathbf{k}}^{\chi}+E_{\mathbf{p}-\mathbf{k}}^{\phi}-E_{\mathbf{p}}^{\phi}\big)t\big]\Big)f(sE^{\chi}_{\mathbf{k}})\Bigg]\: -\:\big(\mathbf{p}\longleftrightarrow \mathbf{p}'\big)^*\Bigg\}
\nonumber\\&\qquad -\:\frac{4m^4}{\mathcal{M}^2}\,\rho(\mathbf{p},\mathbf{p}';t)\sum_{s\,=\,\pm}\int_{\mathbf{x}}\int_{\mathbf{k}_1\mathbf{k}_2}\frac{\sin\big[\big(E_{\mathbf{k}_1}^{\phi}+E_{\mathbf{k}_1-\mathbf{k}_2}^{\phi}+sE_{\mathbf{k}_2}^{\chi}\big)t\big]}{E_{\mathbf{k}_1}^{\phi}+E_{\mathbf{k}_1-\mathbf{k}_2}^{\phi}+sE_{\mathbf{k}_2}^{\chi}}\,\frac{s\big[1+f\big(sE_{\mathbf{k}_2}^{\chi}\big)\big]}{2E_{\mathbf{k}_1}^{\phi}2E_{\mathbf{k}_1-\mathbf{k}_2}^{\phi}2E_{\mathbf{k}_2}^{\chi}}\;,
\end{align}
Here, we have introduced the dummy parameter $s=\pm$ in order to simplify the sum over the two energy flows in the thermal contributions (see, e.g., Ref.~\cite{LeBellac}), making use of the identity in Eq.~\eqref{eq:BEiden}. Specifically, we have written
\begin{equation}
\exp\!\big[-iE_{\mathbf{k}}^{\chi}(t-x^0)\big]+2\cos\big[E_{\mathbf{k}}^{\chi}(t-x^0)\big]f\big(E_{\mathbf{k}}^{\chi}\big)\ =\ \sum_{s\,=\,\pm}s\exp\!\big[-is E_{\mathbf{k}}^{\chi}(t-x^0)\big]\big[1+f\big(sE_{\mathbf{k}}^{\chi}\big)\big]\;.
\end{equation}

Notice that all contributions on the right-hand side of Eq.~\eqref{eq:masterafterint} that arise from nonlocal insertions, i.e.~all but the first term in the second line, vanish identically in the limit $t=0$. This is as one would expect, since the nonlocal insertions contain a residual time integral whose support vanishes in the limit $t\to 0$. In addition, the right-hand side is real in the limit $\mathbf{p}'\to\mathbf{p}$ and consistent with $\rho(\mathbf{p},\mathbf{p}',t)=\rho^*(\mathbf{p}',\mathbf{p},t)$, again as it should be.

The disconnected vacuum diagram of the final line of Eq.~\eqref{eq:masterafterint} vanishes at $t=0$. The $T=0$ ($s=+1$) part also vanishes in the limit $t\to \infty$, since
\begin{equation}
\frac{1}{\pi}\,\frac{\sin\big[\big(E_{\mathbf{k}_1}^{\phi}+E_{\mathbf{k}_1-\mathbf{k}_2}^{\phi}+sE_{\mathbf{k}_2}^{\chi}\big)t\big]}{E_{\mathbf{k}_1}^{\phi}+E_{\mathbf{k}_1-\mathbf{k}_2}^{\phi}+sE_{\mathbf{k}_2}^{\chi}}\ \underset{t\,\to\,\infty}{\longrightarrow}\ \delta\big(E_{\mathbf{k}_1}^{\phi}+E_{\mathbf{k}_1-\mathbf{k}_2}^{\phi}+sE_{\mathbf{k}_2}^{\chi}\big)\;,
\end{equation}
the argument of which is strictly positive. At any finite time $t$, the $T=0$ contribution accounts for particle creation out of the vacuum, as permitted by the uncertainty principle. The $T\neq 0$ part ($s=\pm1$) accounts for particle-number changing interactions with the $\chi$ thermal bath. Note that this is also vanishing in the limit $t\to \infty$ for the present setup, since $M^2<4m^2$ is below threshold (for $s=-1$). As noted above, such number-changing processes would not be present for real atoms.

We recall that we have worked in terms of states with a rescaled mass defined by Eq.~\eqref{eq:mrescale}, which depends on the background value of the chameleon field. If we are sensitive to the absolute value of the mass of the matter field, i.e.~we can predict the phase evolution based on the mass measured in a vanishing ambient value of the chameleon field, then we can capture the leading effect on the dynamics by expanding
\begin{align}
\label{eq:freephaseexp}
E^{\phi}_{\mathbf{p}}\:-\:E^{\phi}_{\mathbf{p}'}\ &=\ \tilde{E}^{\phi}_{\mathbf{p}}\:-\:\tilde{E}^{\phi}_{\mathbf{p}'}\:+\:\frac{\tilde{m}^2\braket{X}}{\mathcal{M}}\bigg(1+\frac{\braket{X}}{\mathcal{M}}\bigg)\Bigg(\frac{1}{\tilde{E}^{\phi}_{\mathbf{p}}}\:-\:\frac{1}{\tilde{E}^{\phi}_{\mathbf{p}'}}\Bigg)\nonumber\\&\qquad-\:\frac{\tilde{m}^4\braket{X}^2}{2\mathcal{M}^2}\Bigg(\frac{1}{(\tilde{E}^{\phi}_{\mathbf{p}})^3}\:-\:\frac{1}{(\tilde{E}^{\phi}_{\mathbf{p}'})^3}\Bigg)\:+\:\mathcal{O}\bigg(\frac{\braket{X}^3}{\mathcal{M}^3}\bigg)\;,
\end{align}
where $\tilde{E}^{\phi}_{\mathbf{p}}=\sqrt{\mathbf{p}^2+\tilde{m}^2}$. A quantitative estimate of the leading effect will be given later in Sec.~\ref{sec:discussion}. In addition, we can expand the term
\begin{align}
&-\:\frac{im^2\lambda\braket{X}}{2\mathcal{M}}\left(\frac{1}{E^{\phi}_{\mathbf{p}}}-\frac{1}{E^{\phi}_{\mathbf{p}'}}\right)\rho(\mathbf{p},\mathbf{p}';t)\Delta^{\rm F}_{xx}\,\frac{\cos(Mt)-1}{M^2}\ =\ -\:\frac{i\tilde{m}^2\lambda\braket{X}}{2\mathcal{M}}\bigg[\bigg(1+\frac{\braket{X}}{\mathcal{M}}\bigg)\Bigg(\frac{1}{\tilde{E}^{\phi}_{\mathbf{p}}}-\frac{1}{\tilde{E}^{\phi}_{\mathbf{p}'}}\Bigg)\nonumber\\&\quad -\:\frac{\tilde{m}^2\braket{X}}{\mathcal{M}}\bigg(1+\frac{\braket{X}}{\mathcal{M}}\bigg)\Bigg(\frac{1}{(\tilde{E}^{\phi}_{\mathbf{p}})^3}-\frac{1}{(\tilde{E}^{\phi}_{\mathbf{p}'})^3}\Bigg)\bigg]\rho(\mathbf{p},\mathbf{p}';t)\Delta^{\rm F}_{xx}\,\frac{\cos(Mt)-1}{M^2}\:+\:\mathcal{O}\bigg(\frac{\braket{X}^3}{\mathcal{M}^3}\bigg)\;.
\end{align}
We remark, however, that a naive expansion about $m^2\sim\tilde{m}^2$ cannot be made in the time-dependent exponentials in the third to seventh lines of Eq.~\eqref{eq:masterafterint}.

\subsection{Renormalization}
\label{sec:Renormalization}

The terms in Eq.~\eqref{eq:masterafterint} yield quadratic and logarithmic ultraviolet divergences. At this order, there are three relevant counterterms: the mass counterterms for the $\phi$ and $\chi$ fields, and the tadpole counterterm for the $\chi$ field.

The tadpole divergences in the second line of Eq.~\eqref{eq:masterafterint} can be renormalized by the standard counterterms, calculated in vacuum. 
On the other hand, the logarithmic divergence arising in the third to fifth lines has acquired a non-trivial time-dependent modulation by virtue of the fact that the interactions have a finite domain of support in time due to the quench, as have the divergences arising from the vacuum diagram in the final line.
 In particular, we see that the contributions vanish identically in the limit $t\to0$. As such, and were we to subtract $t$-independent contributions from the vacuum counterterms, the divergence would persist for all times, except in the limit $t\to \infty$. It follows therefore that the contributions from the relevant counterterms must also vanish in the limit $t \to 0$ and carry the same $t$ dependence.

In order to see this more explicitly, it is instructive to rewrite the divergent terms in the third to fifth lines of Eq.~\eqref{eq:masterafterint} in terms of the more familiar expression for the self-energy. Proceeding in this way and ignoring the thermal corrections (i.e.~the terms which vanish at zero temperature), since these are ultraviolet finite and therefore not relevant to the renormalization, we find that
\begin{align}
\label{eq:beforecounter}
\partial_t\rho(\mathbf{p},\mathbf{p}';t) \ &\supset\ 
\rho(\mathbf{p},\mathbf{p}';t)\Bigg\{\frac{1}{E^{\phi}_{\mathbf{p}}}\int_{p^0}\frac{\sin\big[\big(p^0-E^{\phi}_{\mathbf{p}}\big)t\big]}{p^0-E^{\phi}_{\mathbf{p}}}\,i\Pi^{(T=0)}_{\rm non-loc}(-\,p^2)\:-\:\big(\mathbf{p}\longleftrightarrow \mathbf{p}'\big)^*\Bigg\}\;,
\end{align}
where
\begin{equation}
i\Pi_{\rm non-loc}^{(T=0)}(-\,p^2)\ =\ \bigg(-\frac{2im^2}{\mathcal{M}}\bigg)^2\int_k\frac{-\,i}{k^2+M^2-i\epsilon}\,\frac{-\,i}{(k-p)^2+m^2-i\epsilon}
\end{equation}
is the usual nonlocal, bubble self-energy. (Here, we refer to self-energies that depend on the external momentum flow as nonlocal, and those that are independent of the external momentum flow, i.e.,~tadpoles, as local.) We see that the convolution integral over $p^0$ in Eq. (\ref{eq:beforecounter}) accounts for the finite-time effects, and it must be present also for the counterterm if the divergence is to be removed for all times $t$.
 In particular, it encodes our inability to know precisely the energy scale at which processes are occurring, due to the uncertainty principle, since the external preparation, quench and measurement of the system take place over a finite interval of time.

It follows then that the relevant contribution to the counterterm action must have the same temporal support, i.e.
\begin{equation}
\label{eq:biloccounter}
\delta \widehat{S}_{\rm IF}\ \supset \ -\:\frac{1}{2}\sum_{a\,=\,\pm}a\int_{xy\,\in\,\Omega_t}\delta m^2_{xy}\phi_x^a\phi_y^a\;,
\end{equation}
with $\delta m^2_{xy}$ having the double Fourier transform
\begin{equation}
\delta m^2(p,p')\ =\ \int_x\int_ye^{-ip.x}e^{ip'.y}\delta m^2_{xy}\ =\ (2\pi)^4\delta^4(p-p')\mathrm{Re}\,\Pi^{(T=0)}(-\,p^2)\Big|_{\mathbf{p}\,=\,\bar{\mathbf{p}}}\;,
\end{equation}
which maintains a nontrivial dependence on $p^0$ and therefore $x^0-y^0$. Here,
\begin{equation}
i\Pi^{(T=0)}(-\,p^2)\ =\ i\Pi^{(T=0)}_{\rm loc}\:+\:i\Pi^{(T=0)}_{\rm non-loc}(-\,p^2)
\end{equation}
also contains the local, tadpole self-energy given by
\begin{equation}
i\Pi^{(T=0)}_{\rm loc}\ =\ -\:\frac{2i m^2}{\mathcal{M}^2}\,\Delta^{{\rm F}(T=0)}_{xx}\ = \ -\:\frac{2i m^2}{\mathcal{M}^2}\int_k\frac{-i}{k^2+M^2-i\epsilon}\;.
\end{equation}
Since $i\Pi_{\rm loc}$ is independent of the four-momentum $p$, it yields a time-independent contribution to the quantum master equation. We note that only the dispersive (real) part of the one-loop self-energy is subtracted, since this is where the divergence resides. 
The subtraction point for the renormalization must be taken at a fixed \emph{three}-momentum $\bar{\mathbf{p}}$, since both time-translational and Lorentz invariance are broken by the external maniplulation of the system. One cannot, for instance, straightforwardly apply on-shell renormalization, since the loop corrections to the master equation do not depend only on the Lorentz scalar $p^2$.

The resulting contribution to the master equation from the mass counterterm involves the integral
\begin{align}
\label{eq:counter}
\int_{p^0}\frac{\sin\big[\big(p^0-E^{\phi}_{\mathbf{p}}\big)t\big]}{p^0-E^{\phi}_{\mathbf{p}}}\,\delta m^2(p^0)\;,
\end{align}
where $\delta m^2(p^0)\equiv\int_{p'}\delta m^2(p,p')$, carrying the same modulation as in Eq.~\eqref{eq:beforecounter}. We see that this amounts to a weighted integral over counterterms evaluated at different energy scales, consistently renormalizing the divergences from all competing scales permitted by the uncertainty principle in processes occuring in a finite interval of time. 
Most importantly, in the limit $t\to\infty$, we have
\begin{equation}
\frac{1}{\pi}\,\frac{\sin\big[\big(p^0-E^{\phi}_{\mathbf{p}}\big)t\big]}{p^0-E^{\phi}_{\mathbf{p}}}
\ \underset{t\,\to\,\infty}{\longrightarrow}\ \delta(p^0-E_{\mathbf{p}}^{\phi})\;,
\end{equation}
setting $-\,p^2=m^2$ on-shell, such that we recover the usual on-shell renormalization, with the counterterm fixed at a single energy scale. The $t$-dependent factors appearing throughout should be compared with those that arise in the modified Feynman rules of the interaction-picture formulation of nonequilibrium field theory~\cite{Millington:2012pf, Millington:2013isa}.

The addition of bilocal terms, as in Eq.~\eqref{eq:biloccounter} (that is, terms that depend on two spacetime coordinates), to the action of an open system is not so unusual. Specifically, we can regard the modification of the temporal support of the counterterm above as arising from a correction to the usual bilocal source (see, e.g., Refs.~\cite{Calzetta:1986cq, Berges:2004yj, Millington:2012pf}) that can be used to encode the impact of the environment on the quadratic fluctuations of the open system in approaches based on the (two-particle irreducible) quantum effective action~\cite{Cornwall:1974vz}, as embedded in the Schwinger-Keldysh closed-time-path formalism~\cite{Schwinger:1960qe, Keldysh:1964ud}.

Putting everything together, the full form of the counterterm action (including only the terms relevant at the order we are working) is
\begin{equation}
	\delta \widehat{S}_{\rm IF}\ = \ -\sum_{a\,=\,\pm}a\bigg[\int_{x}\delta\alpha\,\chi^a_x\:+\:\frac{1}{2}\int_{xy}\delta m^2_{xy}\phi_x^a\phi_y^a\:+\:\frac{1}{2}\int_{xy}\delta M^2_{xy}\chi_x^a\chi_y^a\bigg]\;,	
	\label{eq:counterIF}
\end{equation}
with
\begin{subequations}
\begin{align}
\delta m^2_{xy}\ &=\ -\:\frac{2m^2}{\mathcal{M}^2}\,\Delta^{{\rm F}(T=0)}_{xx}\,\delta^{(4)}_{xy}\:+\:\int_{pp'}e^{ip.x-ip'.y}(2\pi)^4\delta^4(p-p')\mathrm{Re}\,\Pi^{(T=0)}_{\rm non-loc}(-\,p^2)\Big|_{\mathbf{p}\,=\,\bar{\mathbf{p}}}\;,\\
\delta M^2_{xy}\ &=\ -\:\bigg[\frac{2m^2}{\mathcal{M}^2}\,D^{{\rm F}(T=0)}_{xx}+\frac{\lambda}{2}\,\Delta^{{\rm F}(T=0)}_{xx}\bigg]\,\delta^{(4)}_{xy}\nonumber\\&\qquad\qquad+\:\int_{pp'}e^{ip.x-ip'.y}(2\pi)^4\delta^4(p-p')\mathrm{Re}\,\Sigma^{(T=0)}_{\rm non-loc}(-\,p^2)\Big|_{\mathbf{p}\,=\,\bar{\mathbf{p}}}\;,\\
\delta \alpha\ &=\ -\:\frac{m^2}{\mathcal{M}}\,D^{{\rm F}(T=0)}_{xx}\:-\:\frac{\lambda\braket{X}}{2}\,\Delta^{{\rm F}(T=0)}_{xx}\;,
\end{align}
\end{subequations}
for any given regularization procedure (e.g.~dimensional regularization). Here, we have introduced the nonlocal chameleon self-energy
\begin{equation}
i\Sigma_{\rm non-loc}(-\,p^2)\ =\ \bigg(-\frac{2im^2}{\mathcal{M}}\bigg)^2\int_k\frac{-\,i}{k^2+m^2-i\epsilon}\,\frac{-\,i}{(k-p)^2+m^2-i\epsilon}\;.
\end{equation}

After making the subtraction, all that remains of the tadpole diagrams in the second line of Eq.~\eqref{eq:masterafterint} are the thermal parts for the $\chi$ field, given by the integral
\begin{equation}
\Delta^{{\rm F}(T\neq 0)}_{xx}\ \equiv\ 2\int{\rm d}\Pi_{\mathbf{k}}\,f\big(E_{\mathbf{k}}^{\chi}\big)\ =\ \frac{T^2}{2\pi^2}\int_{M/T}^{\infty}{\rm d}\xi\,\frac{\sqrt{\xi^2-(M/T)^2}}{e^{\xi}-1}\;,
\label{eq:DeltaF_ren}
\end{equation}
which reduces to $T^2/12$ in the limit $M=0$ and $\sim T^2/29$ for $M/T\sim 1$. The vacuum diagram is renormalized by the two mass counterterms at the level of its two subdiagrams: the nonlocal one-loop $\phi$ and $\chi$ self-energies. Since the loop integrals arising from the nonlocal diagrams cannot be performed in closed form, we do not present these explicitly.


\section{Discussion}
\label{sec:discussion}

The expression \eqref{eq:masterafterint} after the renormalization [see Eq.~\eqref{eq:counterIF}] provides the one-loop master equation describing the open quantum dynamics of a scalar matter field $\phi$ induced by the light scalar $\chi$. To discuss the content and implications of this dynamics, we first rewrite the master equation as
\begin{align}
	\label{eq:masterRen}
	\partial_t\rho(\mathbf{p},\mathbf{p}';t) \ &=\ -\:\big[ i u({\bf p},{\bf p}';t) + \Gamma({\bf p},{\bf p}';t) \big] \rho(\mathbf{p},\mathbf{p}';t)\:+\: \int_{\bf k} \gamma({\bf p},{\bf p}',{\bf k};t) \rho({\bf p-k},{\bf p'-k};t)\;,
\end{align}
where we have defined
\begin{subequations}
	\label{eq:coeffs}
	\begin{align}
		u({\bf p},{\bf p}';t) \ &\equiv\ E^{\phi}_{\mathbf{p}}\:-\:E^{\phi}_{\mathbf{p}'}\:+\:\frac{m^2}{\mathcal{M}}\left(\frac{1}{E^{\phi}_{\mathbf{p}}}-\frac{1}{E^{\phi}_{\mathbf{p}'}}\right)\Delta^{{\rm F}(T\neq 0)}_{xx}\bigg\{
\frac{1}{\mathcal{M}}
+
\frac{\lambda}{2}\braket{X}\frac{\cos(Mt)-1}{M^2}\bigg\}\nonumber\\&\qquad-\,\Bigg\{\frac{1}{E^{\phi}_{\mathbf{p}}}\int_{p^0}\frac{\sin\big[\big(p^0-E^{\phi}_{\mathbf{p}}\big)t\big]}{p^0-E^{\phi}_{\mathbf{p}}}\,\Big[{\rm Re}\,\Pi_{\rm non-loc}(-\,p^2)\nonumber\\&\qquad-\:{\rm Re}\,\Pi^{(T=0)}_{\rm non-loc}(-\,p^2)\big|_{\mathbf{p}\,=\,\bar{\mathbf{p}}}\Big]\: -\:\big(\mathbf{p}\longleftrightarrow \mathbf{p}'\big)\Bigg\}\;, \label{eq:coeffs_u} \\
		\Gamma({\bf p},{\bf p}';t) \ &\equiv\ \frac{1}{E^{\phi}_{\mathbf{p}}}\int_{p^0}\frac{\sin\big[\big(p^0-E^{\phi}_{\mathbf{p}}\big)t\big]}{p^0-E^{\phi}_{\mathbf{p}}}\,{\rm Im}\,\Pi_{\rm non-loc}(-\,p^2)\: +\:\big(\mathbf{p}\longleftrightarrow \mathbf{p}'\big)\;, \label{eq:coeffs_Gamma}\\
		\gamma({\bf p},{\bf p}',{\bf k};t) \ &\equiv \
i\,\frac{4m^4}{\mathcal{M}^2}\sum_{s\,=\,\pm}\Bigg\{
\frac{1}{2E^{\chi}_{\mathbf{k}}2E^{\phi}_{\mathbf{p}-\mathbf{k}} 2E^{\phi}_{\mathbf{p}'-\mathbf{k}}}\,\frac{s}{sE_{\mathbf{k}}^{\chi}+E_{\mathbf{p}-\mathbf{k}}^{\phi}-E_{\mathbf{p}}^{\phi}}\nonumber\\&\qquad\times\Big(1-\exp\big[i(sE_{\mathbf{k}}^{\chi}+E_{\mathbf{p}-\mathbf{k}}^{\phi}-E_{\mathbf{p}}^{\phi}\big)t\big]\Big)f\big(sE_{\mathbf{k}}^{\chi}\big) - ({\bf p} \longleftrightarrow {\bf p}')^*\Bigg\} \;. \label{eq:coeffs_gamma}
	\end{align}
\end{subequations}
We have omitted the contribution from the disconnected vacuum diagram. We note that the nonlocal self-energy $\Pi_{\rm non-loc}(-\,p^2)$ appearing here contains also the thermal corrections, i.e.
\begin{equation}
i\Pi_{\rm non-loc}(-\,p^2)\ =\  \bigg(-\frac{2im^2}{\mathcal{M}}\bigg)^2\int_k\bigg[\frac{-\,i}{k^2+M^2-i\epsilon}\:+\:2\pi f(|k^0|)\delta(k^2+M^2)\bigg]\,\frac{-\,i}{(k-p)^2+m^2-i\epsilon}\;.
\end{equation}

The coefficients $u$ and $\Gamma$ are real, whereas $\gamma$ is complex. This decomposition provides a clear interpretation of the resulting master equation: $u$ corresponds to coherent evolution, resulting from the mass shifts, $\Gamma$ corresponds to decays and, together with the real part of $\gamma$, is responsible for decoherence; $\gamma$ also accounts for momentum diffusion, due to the coupling between the different momentum states. We note that $u({\bf p},{\bf p};t)=0$, as it should, since diagonal elements of the density matrix must be real. 

The master equation \eqref{eq:masterRen} is time local, but with time-dependent coefficients. In general, such master equations do not necessarily preserve the trace and positivity of the density matrix, unless they are in Lindblad form~\cite{Hall_PRA_2014}.\footnote{The possible mapping of time-local master equations to master equations in Lindblad form has been considered in Ref.~\cite{Hush_PRA_2015} by coupling the system described by the non-Lindblad master equation to an ancilla.} 
It is anticipated that any violation of trace preservation and positivity will be of the same order of magnitude as the number-changing processes that have been neglected by restricting to the single-particle subspace for the probe system. In the non-relativistic limit, relevant, for instance, to atomic probes, such processes are expected to be highly suppressed.

Before concluding, we provide an indicative estimate of the order of magnitude of the effects induced by the
light scalar.  We consider the effects on a system of atoms in a chameleon scalar environment in an idealized experimental setup, 
wherein the thermal corrections are assumed to be negligible. 

\begin{figure}[t!]

\centering

\includegraphics[scale=0.7]{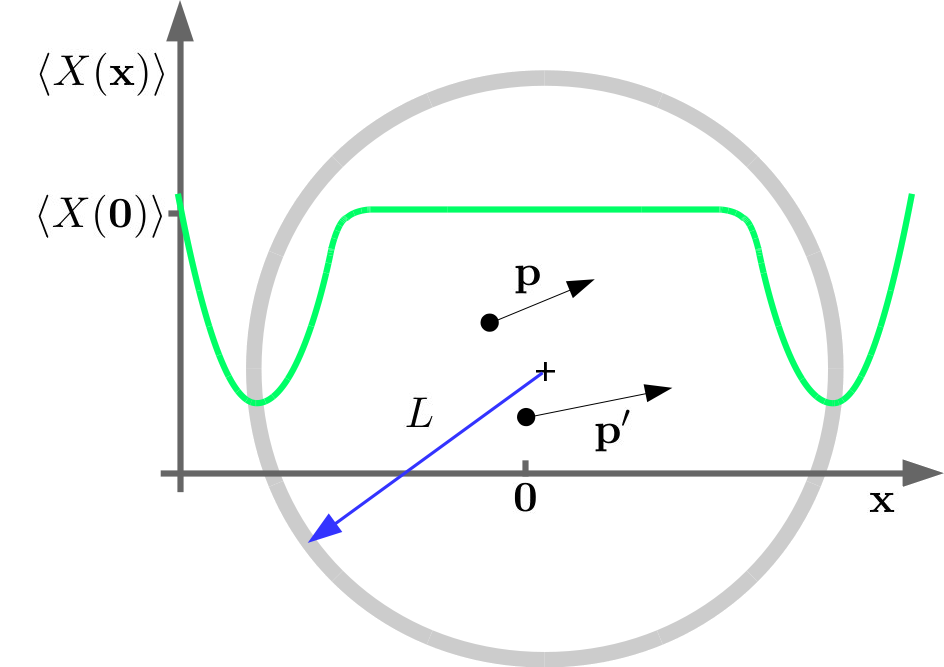}

\caption{\label{fig:chamber} Schematic of the experimental setup. The green line represents the profile of the chameleon field $\braket{X({\bf x})}$ across a cross section of the spherical vacuum chamber of radius $L$ (in grey). At a given time $t$, the test mass is in a superposition of two momenta ${\bf p}$ and ${\bf p}’$ (and in general at different positions as in atom interferometers) and its evolution is described by the master equation (\ref{eq:masterafterint}).}

\end{figure}

Tests of the coherence and decoherence of atomic systems typically take place in high-quality vacuum chambers. If we assume the vacuum chamber is spherical then we can predict the form of the background  chameleon field profile inside the chamber. The walls of the chamber are dense, and so the chameleon is massive, and we can take the chameleon field to be constant in the walls of the chamber.  Inside the vacuum chamber, the chameleon can be much lighter, and the field evolves toward the value that minimizes the effective potential, Eq.~\eqref{eq:V bare cham}, when $\rho^{\rm ext}$ is the density of the residual gas in the chamber. However, over a large part of the chameleon parameter space, and for typical vacuum chamber sizes, there is not enough space for the chameleon to reach the minimum of the effective potential.  In this case, the field adjusts its value so that its Compton wavelength becomes of order the size of the vacuum chamber, and, at the center of the spherical chamber, the expectation value of the chameleon field and the mass of its fluctuations are given by~\cite{Burrage:2014oza}
\begin{align}
\braket{X} \ =\  -\frac{q}{\sqrt{\lambda}L}\;,\qquad M \ =\  \frac{q}{\sqrt{2}L}\;,
\end{align}
where $q= 1.287$~\cite{Khoury:2003rn, Burrage:2014oza} and $L$ is the radius of a spherical vacuum chamber.  The form of the chameleon profile inside the chamber is illustrated schematically in Fig.~\ref{fig:chamber}.

In experiments with microscopic test masses, one will typically be in the regime \mbox{$\tilde{m}/\mathcal{M} \ll1$}. 
Therefore, in order to provide a conservative upper estimate on the order of magnitude of the effects, we assume that we are sensitive to the dominant coherent shift arising from the change in the effective mass of the matter field due to the background value of the chameleon. We consider the leading contribution in the first power of $\tilde{m}/\mathcal{M}$, and ignore the subdominant thermal corrections. As identified in Sec.~\ref{sec:qme}, the relativistic loop corrections obtained for the toy scalar model used here to facilitate the main formal developments are not expected to be reliable proxies for diagrams involving nonrelativistic and extended probes, such as atoms, and we therefore refrain from using these for the indicative estimate that follows. The application of the present formalism to more realistic probe models will be presented elsewhere.

Motivated primarily by using precision atom-interferometry measurements, we rewrite the free-phase part of Eq.~\eqref{eq:coeffs_u} via Eq.~\eqref{eq:freephaseexp}, performing the non-relativistic expansion of the $1/\tilde{E}$ terms to find
\begin{equation}
	|\Delta u| \ \approx\ \left|\frac{\braket{X}}{2\mathcal{M}}\,\tilde{m}\bigg[1\:-\:\frac{\lambda T^2}{29M^2}\bigg]\frac{v^2}{c}\right|\;.
	\label{eq:coeffs_u_est}
\end{equation}
Here, we have defined the characteristic velocity scale by $v^2 = \big||{\bf p}|^2 - |{\bf p}'|^2\big|/\tilde{m}^2$, taken the maximum value of $1-\cos(Mt)=2$, used the relation \eqref{eq:DeltaF_ren} for $\Delta_{xx}^{{\rm F}(T\neq 0)}$, and restored the dimensions, so that $[u] = {\rm Hz}$ for $T$ expressed in mass units and $\braket{X}$ expressed in units of inverse length, as it should for a rate of change. The thermal correction, included  here for illustration, corresponds to the thermal shift in the background chameleon field to smaller values, arising from the third term in the second line of Eq.~\eqref{eq:masterafterint}.

As a concrete example, we consider the quantum test mass to be the ${}^{87}{\rm Rb}$ isotope with $\tilde{m} = 87 m_u$ (where $m_u$ is an atomic mass unit) and choose $\lambda = 1/10$ and $\mathcal{M} = M_\text{Pl}$, $L = 1\ \text{m}$ for the radius of the vacuum chamber and $v =10\ {\rm m}\,{\rm s}^{-1}$ (velocities of up to $6\ {\rm m}\,{\rm s}^{-1}$ were reported in atomic transport experiments~\cite{Schmid_2006}) for the characteristic velocity scale. We further consider the vacuum chamber to be in a thermal equilibrium at temperature $T=1\;{\rm mK}$.  The chosen values imply $M \approx 10^{-16}m_u$ and $M/T \sim 1$,\footnote{Noticing that we are in a regime $T\sim M$, we might be concerned about non-perturbative effects arising from the thermal corrections to the dynamics of the chameleon field itself. A comprehensive study of this is beyond the scope of the present work.} in which case the thermal shift to the chameleon background field value is subleading. We then obtain $|\Delta u| \approx 10^{-23}$~Hz, which is far out of reach of the current atom-interferometry sensitivity of order $10^{-8}$~Hz.\footnote{This value is inferred from Ref.~\cite{Estey_2015}, which reported a phase measurement with statistical uncertainty of $10^{-8}$ rad obtained after $\sim 1$ day of integration time and for a duration of the order of 1 s for each experimental run.}

The conservative value of $\mathcal{M} = M_\text{Pl}$ is motivated by constraints on chameleon theories~\cite{Burrage:2016bwy}, and the value of $\lambda < 1$ was chosen to remain well within the regime of perturbative validity. We reiterate, however, that the chosen parameters are in tension with the current experimental bounds for the chameleon potential considered here, as discussed earlier in Sec.~\ref{sec:Chameleon}. Even so, our aim is to provide a conservative estimate of the order of magnitude of the effects. We see that the effects induced by the quartic chameleon considered here are negligibly small and potentially out of reach of any near-future experiment. This is consistent with the fact that the classical fifth force, i.e.~which also depends on the classical background field value $\braket{X}$, is constrained to be very small.  

It therefore seems that, while atom interferometry provides a powerful tool in the search for fifth forces due to light scalar fields, a direct detection of the effects induced by the quantum fluctuations of the light fields remains extremely challenging. Here, it would be interesting to apply the present theory to the other light scalar field models discussed in Sec.~\ref{sec:lightscalar} in order to obtain quantitative estimates. While it is likely that the induced effects will remain elusive also in these cases, we note that possible pathways for improvements include using more massive test masses to increase the ratio $\tilde{m}/\mathcal{M}$, provided they do not affect the screening mechanism. In this context, we note that a Bell test was recently performed with levitated nanoparticles containing $10^{10}$ silica atoms~\cite{Marinkovic_PRL_2018}, amounting to an increase of $\tilde{m}$ by 10 orders of magnitude as compared to single-atom interferometry. Other possibilities include, for example, considering the effects induced in other detection platforms, in particular in optical atomic clocks, which provide the most accurate measurement tool currently being developed, with reported relative precision reaching $10^{-19}$~\cite{Marti_PRL_2018}. Alternatively, one could consider modifying the experimental setup so that each branch of the interferometer experiences a different background value of the light field together with its gradient. In such a case, the description provided here [Eq.~\eqref{eq:masterRen}] has to be extended so as to account for the spatially inhomogeneous background provided by the screening field. We leave this and related extensions for future work.

Finally, we would like to comment on the difference of the current analysis with that of Ref.~\cite{Burrage:2014oza}, where a proposal of testing the modification of gravitational acceleration $g \rightarrow g + \delta g$ induced by the chameleon scalar fields has been put forward. Specifically, the change $\delta g$ predicted in Ref.~\cite{Burrage:2014oza} scales as $\delta g \propto \partial_X V(X)$, i.e., with the \emph{gradient} of the chameleon potential. This is a classical effect that arises because the length of the paths that the atoms explore in the interferometer depends on the fifth force due to the gradients of the background chameleon field configuration. This is not the same experimental setup as considered in this work, since, in order to detect the classical fifth force due to the chameleon, a macroscopic source mass must also be placed inside the vacuum chamber. For the choices of the chameleon potential $\lambda \approx 1/10$ and $\mathcal{M} \sim M_{\rm Pl}$, and a vacuum chamber similar to the simplified chamber we consider here, a phase shift of order $10^{-3}$ would be induced if the interferometry experiment were performed with rubidium atoms held within 1 cm of a massive sphere of radius 1 cm and density 1 ${\rm g}\,{\rm cm}^{-3}$. This should be contrasted with the prediction of Eq.~\eqref{eq:coeffs_u} for the coherence shift, which scales with the background field value instead, i.e.~$u \propto \braket{X}$; see also~Fig.~\ref{fig:potential}.

\section{Conclusions and outlook}
\label{sec:conc}

Light scalar fields, which couple to matter either conformally or through a Higgs portal, are well motivated as extensions of the Standard Model and/or general relativity. Models such as the chameleon, discussed here, possess a screening mechanism that allows them to avoid local searches for fifth forces while remaining light on cosmological scales and coupling to matter with at least gravitational strength. In this work, we have developed, from first principles, a novel approach for describing how an environment composed of such a scalar field affects the dynamics of a probe field. This is, to the best of our knowledge, the first work in which the quantum dynamics of both a probe field and the (conformally coupled) light scalar are studied; excepting~Ref.~\cite{Brax:2018grq}, previous work looking at the effects of screened scalar fields in, for example, atom-interferometry experiments treated the scalar only classically.

Herein, we have used a second scalar field as a proxy for the quantum probe, imagining, for instance, an atom in an atom-interferometry experiment. While this simple scalar field theory provides a convenient playground in which to develop the necessary techniques, it nevertheless has some shortcomings as a toy model of an atom, such as allowing for physically unrealistic decay and production processes, and the extension of this work to more realistic models of probe systems may be presented elsewhere.

Beginning from the path-integral approach of the Feynman-Vernon influence functional, we have employed an LSZ-like reduction technique, constructed in the operator-based framework of thermo field dynamics, which allows us to make use of diagrammatic techniques from (nonequilibrium) quantum field theory, while also making a concrete connection with the single-particle matrix elements of the reduced density operator of interest for quantum probes. In addition, we have shown that the consistent renormalization of the resulting master equation in the non-Markovian, postquench regime requires the introduction of time-modulated counterterms, allowing us to make quantitative predictions that are independent of any ultraviolet cutoff.

The final master equation describes the coherent dynamics of the probe field, as well as decoherence and momentum diffusion. Having access to quantitative estimates of these effects, we have confirmed that their experimental observation remains a challenge. Even so, the present formalism allows us to identify possible pathways for improvements in future searches, including the use of more massive probe fields and the studies of the effects in optical atomic clocks or nonhomogeneous scalar-field backgrounds. Importantly, the present work provides a robust and complementary approach for studying open quantum dynamics, which may shed new light on, e.g., the contentious area of gravitational decoherence.


\begin{acknowledgments}

We thank E.~Aurell, T.~Fernholz, P.~Haslinger, B.-L.~Hu, F.~C.~Lombardo, M.~Marcuzzi, T.~Prokopec, G.~Tasinato, and Y.~Zhang for useful comments and illuminating discussions. P.M. thanks A.~Pilaftsis for earlier collaboration on similar topics, as well as R.~Dickinson and J.~Forshaw for many discussions on the operator-based approach of thermo field dynamics. C.B. is supported by a Royal Society University Research Fellowship; both C.B. and P.M. are supported by a Leverhulme Trust Research Leadership Award. P.M. acknowledges the partial support of STFC Grant No.~ST/L000393/1. C.K. is supported by the School of Physics and Astronomy of the University of Nottingham and via a Vice-Chancellor's Scholarship for Research Excellence. J.M. acknowledges support from the European Research Council under the ERC Grant Agreement No.~335266 (ESCQUMA) and from the EPSRC Grant No. EP/P026133/1.

\end{acknowledgments}


\bibliography{open_dynamics}
\bibliographystyle{JHEP}

\end{document}